\documentclass[aps,prb,showpacs,superscriptaddress,twocolumn]{revtex4-2}

\usepackage{amsmath,graphics}
\usepackage[next]{inputenc}
\usepackage[dvips]{epsfig}
\usepackage[colorlinks=true,citecolor=blue,linkcolor=blue]{hyperref}
\usepackage{bbm}
\usepackage{bbold}
\usepackage{booktabs}
\usepackage{multirow}
\usepackage{hhline}
\usepackage{graphicx}
\usepackage{amsmath}
\usepackage{multirow}
\usepackage{lipsum}

\begin{document}

\title{Topological Nernst and topological thermal Hall effect in rare-earth kagome ScMn$_6$Sn$_6$}
\author{Richa P. Madhogaria}
    \email{rpokhar1@utk.edu}
    \affiliation{Department of Materials Sciences and Engineering, University of Tennessee, Knoxville, TN 37996, USA}
    \author{Shirin Mozaffari}
    \affiliation{Department of Materials Sciences and Engineering, University of Tennessee, Knoxville, TN 37996, USA}
    \author{Heda Zhang}
    \affiliation{Materials Science and Technology Division, Oak Ridge National Laboratory, Oak Ridge, Tennessee 37831, USA}
    \author{William R. Meier}
    \affiliation{Department of Materials Sciences and Engineering, University of Tennessee, Knoxville, TN 37996, USA}
    \author{Seung-Hwan Do}
    \affiliation{Department of Materials Sciences and Engineering, University of Tennessee, Knoxville, TN 37996, USA}
    \author{Rui Xue}
    \affiliation{Department of Physics and Astronomy, University of Tennessee Knoxville, Knoxville, TN 37996, USA}
    %\affiliation{Department of Physics and Astronomy, University of Tennessee Knoxville, Knoxville, TN 37996, USA}   
    \author{Takahiro Matsuoka}
    \affiliation{Department of Materials Sciences and Engineering, University of Tennessee, Knoxville, TN 37996, USA}
    \author{David G. Mandrus}
    \email{dmandrus@utk.edu}
    \affiliation{Department of Materials Sciences and Engineering, University of Tennessee, Knoxville, TN 37996, USA}
    \affiliation{Materials Science and Technology Division, Oak Ridge National Laboratory, Oak Ridge, Tennessee 37831, USA}
    \affiliation{Department of Physics and Astronomy, University of Tennessee Knoxville, Knoxville, TN 37996, USA}

\begin{abstract}
 Thermal and thermoelectric measurements are known as powerful tools to uncover the physical properties of quantum materials due to their sensitivity towards the scattering and chirality of heat carriers. We use these techniques to confirm the presence of momentum and real-space topology in ScMn$_6$Sn$_6$. 
 There is an unconventional dramatic increase in the Seebeck coefficient on entering the transverse conical spiral (TCS) below $T$ = 200 K suggesting an unusual scattering of heat carriers.
 %The unconventional change in the percentage of Seebeck coefficient below $T$ = 200 K in the transverse conical spiral (TCS) phase of ScMn$_6$Sn$_6$ reveals the presence of unusual scattering phenomenon of the heat carriers. 
 In addition, the observed anomalous thermal Hall effect and the anomalous Nernst effect indicates non-zero Berry curvature in $k$-space. 
 Furthermore, we identify a significant topological contribution to the thermal Hall and Nernst signals in the TCS phase revealing the impacts of real-space Berry curvature. 
 %Topological contributions to the thermal Hall and Nernst thermopower was also discerned in the TCS phase. 
 We discuss the presence of topological thermal Hall effect and topological Nernst effect for the first time in the diverse HfFe$_6$Ge$_6$ family. 
 %This allows for exploring these effects in systems with non-coplanar spin structures and also motivates to understand the underlying physics behind the origin of geometrical effects in 166 family. 
 This study illustrates the importance of transverse thermal and thermoelectric measurements to investigate the origin of topological transport in the non-coplanar magnetic phases in this family of kagome metals.
\end{abstract}
\date{\today}

\maketitle
\section{Introduction \label{intro}}
Thermal and thermoelectric transport measurements is a rising powerful technique to help discover the intriguing physical properties of quantum materials~\cite{largeANE-AFM,TNE-Nd3Ru4Al12,Mn3Ge_ANE,usedforintro}. Importantly, studying the thermal properties give insight into charge neutral excitations like lattice vibrations (phonons), and magnons, which are not detected via electrical methods. These measurements are compelling for insulators to retrieve information regarding the dynamics of the charge neutral heat carriers and can be equally used for metallic systems~\cite{thermal-hall-insulators,thermal-hall-insulators2,thermal-hall-insulators3,usedforintro,Fe3Sn2}. The transverse thermal and thermoelectric counterparts of the electrical anomalous Hall effect (AHE), $\sigma_{xy}^A$, are anomalous thermal Hall effect (ATHE), $\kappa_{xy}^A$, and anomalous Nernst effect (ANE), S$_{xy}^A$, respectively. As a general rule, the Hall signal can be separated into three components based on their origin: ordinary contributions arising from deflection of carriers by the applied field, anomalous contributions stem from Berry curvature in $k$-space, and a topological contribution from real-space Berry curvature from spin textures. Out of the three, the anomalous parts can be      
mathematically expressed as~\cite{AHE-formula-MacDonald,Mn3Ge_ANE,ANE-formula-UCo0.8Ru0.2Al,Co2MnGa,signofNenst}: 
\begin{equation}\label{AHE}
  \sigma_{xy}^A= -2\pi e^2/h \int_{BZ} (d^3 k)/(2\pi)^3  f(k)\Omega_B (k)
\end{equation}
\begin{equation}\label{ATHE}
  \kappa_{xy}^A= -2\pi /hT \int_{BZ} (d^3 k)/(2\pi)^3  \epsilon f(k)\Omega_B (k)
\end{equation}
\begin{equation}\label{thermoelectric}
  \alpha_{xy}^A= -2\pi e^2/h \int_{BZ} (d^3 k)/(2\pi)^3  \gamma(k)\Omega_B (k)
\end{equation}
\begin{equation}\label{ANE}
   S_{xy} = (\sigma_{xx}\alpha_{xy}^A- \sigma_{xy}^A\alpha_{xx})
  \end{equation}
where $\Omega_{B}$ (k): Berry phase curvature, $f(k)$: Fermi Dirac distribution function, $\epsilon$: energy, $\alpha_{xy}$: the transverse component of the thermoelectric tensor and  $\gamma$(k)= -$f(k)$$\ln$($f(k)$)-(1-$f(k)$) $\ln$(1-$f(k)$)~\cite{ANE-formula-UCo0.8Ru0.2Al}. 
From the mathematical formulation, it is clear that similar to AHE, both ATHE and ANE are also dependent on the Berry curvature (BC) of the electrons bands and hence can serve as two other transport tools to probe the non-trivial BC. All three effects appear when the charge carriers acquire an anomalous transverse velocity and a finite Berry phase in the presence of longitudinal current or thermal gradient~\cite{AHE-formula-MacDonald,largeANE-AFM,TNE-Nd3Ru4Al12}. However, the three anomalous Hall transports provide access to BC with different weights; $f(k)$, $\epsilon f(k)$, and $\gamma$(k) for $\sigma_{xy}^A$, $\kappa_{xy}^A$ and S$_{xy}^A$, respectively. As a consequence, $\kappa_{xy}^A$ and S$_{xy}^A$ provide additional information about the entropy and the velocities of the electrons near the Fermi surface.
\par The third component which is the manifestation of the spin textures, namely, topological  Hall effect (THE) (electrical), topological thermal Hall effect (TTHE) (thermal), and topological Nernst effect (TNE) (thermoelectric) is found to be superimposed on the anomalous signals and most of the time are very subtle to detect. Specifically, these effects have been assigned to the presence of non-zero static spin chirality generated due to the chiral (or noncoplanar) spin texture~\cite{THE-MnSi,Fe3Sn2,GaV4Se8}. Although they share common origin it is not yet well understood why all the materials that exhibit THE doesn't show TTHE and TNE.
\par THE has been widely studied in different classes of materials, from manganites~\cite{THE_manganites} to frustrated magnets~\cite{THE-frustratedmagnet} to the bulk skyrmions~\cite{THE-MnSi}, whereas TNE and TTHE have been extremely difficult to observe in the experiments. While there is only one experimental report for TTHE (GaV$_4$Se$_8$)~\cite{GaV4Se8}, a few materials are reported to host TNE, for example: Fe$_3$Sn$_2$~\cite{Fe3Sn2}, MnGe~\cite{TNE-MnGe}, Gd$_2$PdSi$_3$~\cite{TNE-Gd2PdSi3}, Nd$_3$Ru$_4$Al$_{12}$~\cite{TNE-Nd3Ru4Al12}. The scalar spin chirality coupled with the conduction electrons is responsible for the TNE. The first three aforementioned materials host skyrmions which generate scalar spin chirality, while the spin canting due to the thermal fluctuations at the vicinity of the ferromagnetic  ordering leads to the chiral magnetic state in the case of Nd$_3$Ru$_4$Al$_{12}$. The study on  Nd$_3$Ru$_4$Al$_{12}$ opens up a new perspective to explore the topological Nernst effect beyond skyrmionic systems. The lack of experiments to explore topological effects other than THE motivates us to investigate TTHE and TNE in ScMn$_6$Sn$_6$, a non-skyrmion system.
\par The diverse family of hexagonal HfFe$_6$Ge$_6$ materials provide a huge platform for new phenomena owing to their exotic band structure, which includes Dirac points, van Hove singularities, and flat bands, and their high tunability~\cite{YMn6Sn6-dirac,YMn6Sn6-Nirmal,TbMn6Sn6-cherngap,GaneshPRL,ScV6Sn6}. Room temperature topological Hall effect in YMn$_6$Sn$_6$~\cite{YMn6Sn6-Nirmal}, Chern gap coupled with magnetism and skyrmion spin-lattice in TbMn$_6$Sn$_6$~\cite{TbMn6Sn6-cherngap}, quantum oscillations from non-trivial Berry phase in $R$Mn$_6$Sn$_6$ ($R$ = Gd-Tm,Lu)~\cite{QO-RMn6Sn6}, exotic phonon-driven charge density wave in ScV$_6$Sn$_6$~\cite{ScV6Sn6}, Z$_2$ topological kagome metals~\cite{Ganesh-Z2} are some of the examples which necessitates the study of this class of materials.
\par
\begin{figure}[h]
	\includegraphics[width =8.6 cm]{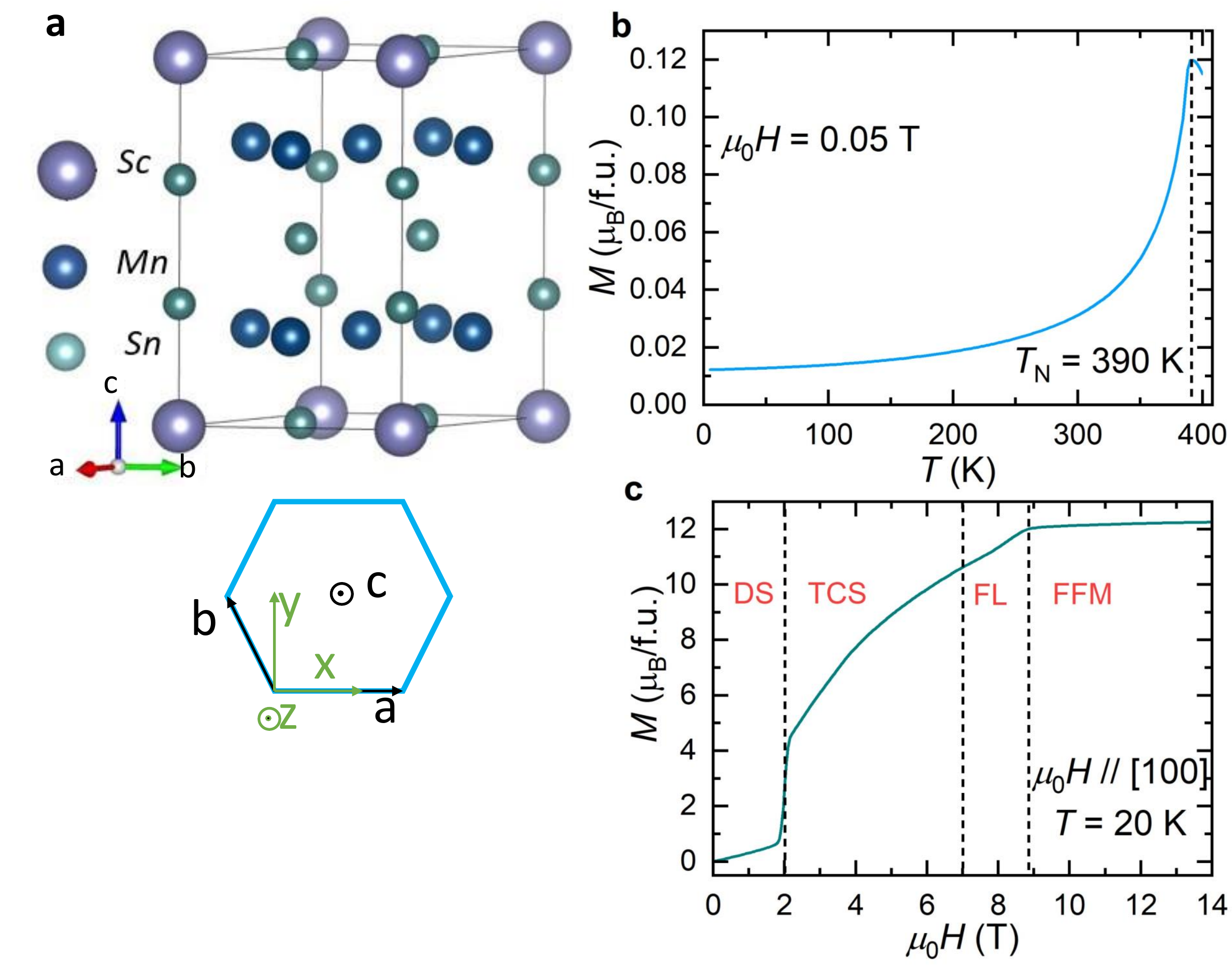}
	\caption{(a) Schematic crystal structure of ScMn$_6$Sn$_6$: the formation of kagome layers by Mn atoms (blue spheres) can be clearly seen (top); (bottom) illustrates the crystal axes and the co-ordinate axes of a hexagonal shaped crystal.(b) Temperature dependent of zero-field cooled magnetization ($M$) curve with. (c) Field dependent magnetization at $T$ = 20 K. For both the magnetization curves the external magnetic field was parallel to [100]. The evolution of different magnetic phases, double spiral (DS) to transverse conical spiral (TCS) to fan-like (FL) to forced ferromagnetic (FFM) can be seen from $M$ vs. $\mu_0H$ graph. 
} \label{Fig1}
\end{figure}

 Figure \Ref{Fig1} (a) (top) shows the schematic crystal structure of ScMn$_6$Sn$_6$. In this material Mn atoms form the double layered kagome sheets which is represented by the blue atoms in the figure. The magnetic spins align ferromagnetically within the kagome sheets and are helically coupled along the $c$-axis~\cite{ScMn6Sn6-APL}. ScMn$_6$Sn$_6$ orders antiferromagnetically below $T_N$ = 390 K [Figure~\Ref{Fig1} (b)] and transitions through multiple magnetic phases as the strength of external magnetic field is increased [Figure~\Ref{Fig1} (c)]. As can be seen from Figure~\Ref{Fig1} (c), the magnetic structure evolves from double spiral (DS) to transverse conical spiral (TCS) to fan-like (FL) and finally to a forced ferromagnetic (FFM) state at high magnetic field~\cite{ScMn6Sn6-APL}. These phases of ScMn$_6$Sn$_6$ are very similar to the one reported for YMn$_6$Sn$_6$~\cite{YMn6Sn6-Nirmal}.
\par In this study we focus to understand how the presence of exotic magnetic phases can affect its transport properties. Here we report a comprehensive thermal and thermoelectric transport study on the kagome metal ScMn$_6$Sn$_6$. Both ATHE and ANE appears at $T$ = 300 K and persists down to the lowest measured temperature, $\it{T}$ = 120 K. The presence of ATHE and ANE manifest the existence of large Berry curvature in ScMn$_6$Sn$_6$. A large anomalous Nernst signal, 2.21 $\mu$VK$^{-1}$, is seen  at room temperature which is similar to that reported for the sister compounds YMn$_6$Sn$_6$~\cite{YMn6Sn6_Nernst} and TbMn$_6$Sn$_6$~\cite{TbMn6Sn6-2ndthermaltransport,TbMn6Sn6-thermaltransport}, and comparable to the highest ANE observed~\cite{Co2MnGa}. In addition to the ATHE and ANE, ScMn$_6$Sn$_6$ also exhibits TNE from $\it{T}$ = 120$-$300 K and TTHE from $\it{T}$ = 120$-$280 K within the field regime $\mu$$_0$$\it{H}$ = 1.6$-$5.3 T. The width of the magnetic field which shows the TNE and TTHE corresponds to the TCS magnetic phase of ScMn$_6$Sn$_6$, revealing a close association between the non-coplanar TCS phase (real-space topology) and the topological effects. Our results establishes the importance of the thermal/thermoelectric measurements to reveal the inherent topology of the quantum materials.

\section{Experiments}\label{Experiments}
\textbf{Growth and characterization}: Single crystals of ScMn$_6$Sn$_6$ were grown via self-flux method. Sc(Alfa Aser 99.9 \%), and Mn(Alfa Aser 99.98\%)
pieces were combined with Sn shot (Alfa Aser 99.999\%) in the atomic ratio Sc:Mn:Sn = 1 : 6 : 30 in a Canfield crucible set~\cite{canfield}. The crucible was then sealed in a silica tube under vacuum. The ampule was heated at $\mathrm{60 ^{\circ}C~h^{-1}}$ to $\mathrm{973 ^{\circ}C}$, followed by a 48 hour dwell.  Crystals were grown during slow cool ($\mathrm{1.2 ^{\circ}C~h^{-1}}$) to $\mathrm{600 ^{\circ}C}$. The hexagonal shaped ScMn$_6$Sn$_6$ crystals were retrieved from the molten Sn flux by centrifuging. The HfFe$_6$Ge$_6$-type structure was confirmed through 
powder X-ray diffraction (PXRD) using a PANalytical Empyrean diffractometer with a Cu-K$_{\alpha}$ source. The chemical composition and homogeneity of the crystals were confirmed through Energy-dispersive X-ray spectroscopy. Fullprof was used for the Reitveld refinement of the room temperature PXRD data [Supplementary Figure S1] and the obtained lattice parameters were $a$ = 5.46493(4) \AA\ and $c$ = 8.96077(4) \AA.
\par \textbf{Magnetization measurements}: Magnetization measurements were carried out using the vibrating sample magnetometer (VSM) option of a physical property measurement system (PPMS) from Quantum Design.
\par \textbf{Transport measurements}: Single crystals were shaped into a rectangular bar such that the external applied magnetic field is parallel to [100] and current/temperature gradient is parallel to [001] crystallographic axes. The electrical measurements were done using the resistivity option of the PPMS. The thermoelectric measurements were performed using a PPMS compatible self-designed sample puck and a break out box. Type E Chromel-Constantan thermocouples were used to measure the temperature gradient. The schematic of the device is shown in Figure S2 of the supplementary. The temperature gradient over the sample was created by attaching one end to the heater with 1 k$\Omega$ resistance and other end to a piece of copper. Keithely current source 2450 was used for applying the current and Keithely 2182A nanovoltmeters were used to record the induced voltage. The longitudinal thermal resistivity (anomalous thermal Hall resistivity) were obtained by symmetrizing (antisymmetrizing) the measured temperature gradient in the presence of positive and negative magnetic field separately. Similarly, Seebeck and Nernst signals were also acquired through symmetrized and antisymmetrized thermoelectric voltages, respectively. The sign of the transverse components have been explained in the supplementary information.

\begin{figure}[h]
	\includegraphics[width =8.6 cm]{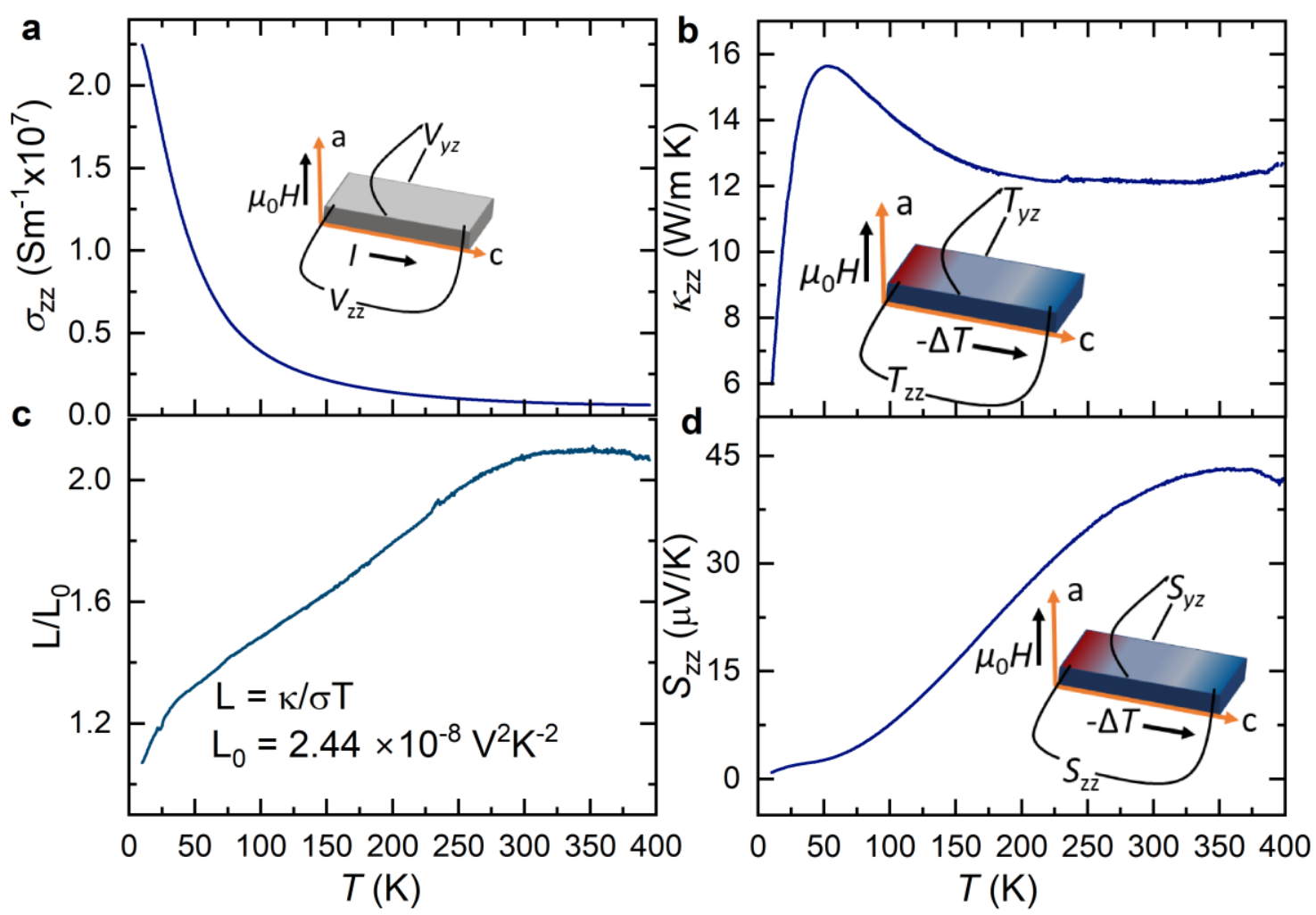}
	\caption{Temperature dependence of electrical and thermal measurements. Electrical conductivity versus temperature (a), thermal conductivity versus temperature (b). (c) Deviation of the Lorenz number from the expected theoretical value with respect to the temperature. (d) Temperature dependence of Seebeck coefficient. The direction of the applied magnetic field  and current/temperature gradient are shown in insets of (a), (b) and (d). The color gradient represent the flow of heat from hot to cold end.
} \label{Fig2}
\end{figure}

\section{Results and Discussions}\label{Results}
We present the bulk electronic and thermal properties of ScMn$_6$Sn$_6$. The temperature dependence of electrical conductivity ($\sigma_{zz}$), thermal conductivity ($\kappa_{zz}$) and Seebeck coefficient ($S_{zz}$)  of ScMn$_6$Sn$_6$ are shown in Figures~\Ref{Fig2} (a), (b) and (d), respectively. The insets of these figures represent the geometry of the experiments used for this study. For electrical measurements as shown in the inset of Figure~\Ref{Fig2} (a), the applied current is parallel to the $z$-axis, magnetic field is along the $x$-axis and the transverse voltage is parallel to the $y$-axis. Similarly, for the thermal and thermoelectric measurements the insets of~\Ref{Fig2}(b) and (d) show the direction of the applied temperature gradient ($||$ $z$-axis), magnetic field ($||$ $x$-axis) and transverse temperature gradient/thermoelectric voltage ($||$ $y$-axis) [ Note: Refer to the bottom image of Figure~\Ref{Fig1} (a) for the crystallographic and co-ordinate axes of hexagonal shaped crystals]. The temperature dependent $\sigma_{zz}$ decreases monotonically [Figure~\Ref{Fig2} (a)] as we warm up from $T$ = 2 $-$ 400 K indicating the metallic nature of the sample. The temperature profile of $\kappa_{zz}$ [Figure~\Ref{Fig2} (b)] reflects a similar trend of the thermal conductivity as expected by the Debye approximation. As the temperature is raised from $T$ = 2 $-$ 400 K, it shows an increase, with a  broad characteristic peak at $T$ = 52.5 K, and then decrease followed with almost saturated value above $T$= 200 K. The peak at $T$ = 52.5 K corresponds to the competition between two phonon scattering mechanisms; low temperature: the phonon scattering being dominated by the defects/boundary, high temperature: this process is governed by the Umklapp phonon scattering~\cite{book_thermal}. Wiedemann-Franz law, ${L}= (\dfrac{\boldsymbol{\kappa}}{T\boldsymbol{\sigma}} )$, states that the ratio of thermal conductivity to the electrical conductivity is temperature independent and can be represented in terms of Lorenz number for a metallic system.  Nevertheless, as shown in Figure~\Ref{Fig2}(c), this number varies with temperature for ScMn$_6$Sn$_6$. This variation from the required value of the $L_0$ = 2.44 $\times$10$^{-8}$V$^2$K$^{-2}$ is suggestive of inelastic phonon scattering~\cite{ashcroft1976solid,Lorenz-number-citation}.
\par Figure~\Ref{Fig2}(d) displays the temperature dependent profile for $S_{zz}$. Seebeck coefficient is defined as $S_{zz}$= -$\partial_zU$/$\partial_zT$, where $U$ is the longitudinal thermoelectric voltage and $T$ is the measured temperature. Alternatively, $S_{zz}$ can also be understood as an averaged entropy flow per charge carrier~\cite{Onsager-reltion}. It can be seen that the magnitude of $S_{zz}$ increases constantly with the increase in temperature and remains positive throughout the measured temperature regime [Figure~\Ref{Fig2}(d)]. The positive value of $S_{zz}$ can be associated with the predominant p-type charge carriers present in the system. Moreover, it has a high magnitude of 40$\mu$VK$^{-1}$ at $T$ = 300 K. To understand what this means, let us look at the Seebeck coefficient based on the Mott relation ~\cite{mahan1979good,MnGe-thermopower},
\begin{equation}\label{Mott-reln}
   S = -\dfrac{\pi^2k_B^2T}{3e}\left[\dfrac{\partial \ln{D(\epsilon)} }{\partial \epsilon} + \dfrac{\partial \ln{\tau(\epsilon)} }{\partial \epsilon}\right]_{\epsilon = \epsilon_F}
  \end{equation}
where $k_B$, $e$, $D(\epsilon)$, $\tau(\epsilon)$, and $\epsilon_F$  are the Boltzmann constant, electronic charge, density of states, relaxation time and Fermi energy, respectively. According to the above relation, for a metallic system $S$ is a few $\mu$VK$^{-1}$ owing to the symmetric band structure around the Fermi level which allows for the cancellation of entropy flow of electrons above and below the Fermi level. Whereas, asymmetric band structure leading to asymmetric density of states contributes to the higher value of $S$~\cite{HeuslerhighS,MnGe-thermopower}. Hence, we suspect that the magnitude of $S_{zz}$  at $T$ = 300 K (40$\mu$VK$^{-1}$) is due to the the nontrivial electronic topology of ScMn$_6$Sn$_6$.
The slight downturn seen in the data at $T$ = 374 K could be due to the proximity of the antiferromagnetic transition.
\begin{figure}[hbt]
	\includegraphics[width=1\linewidth]{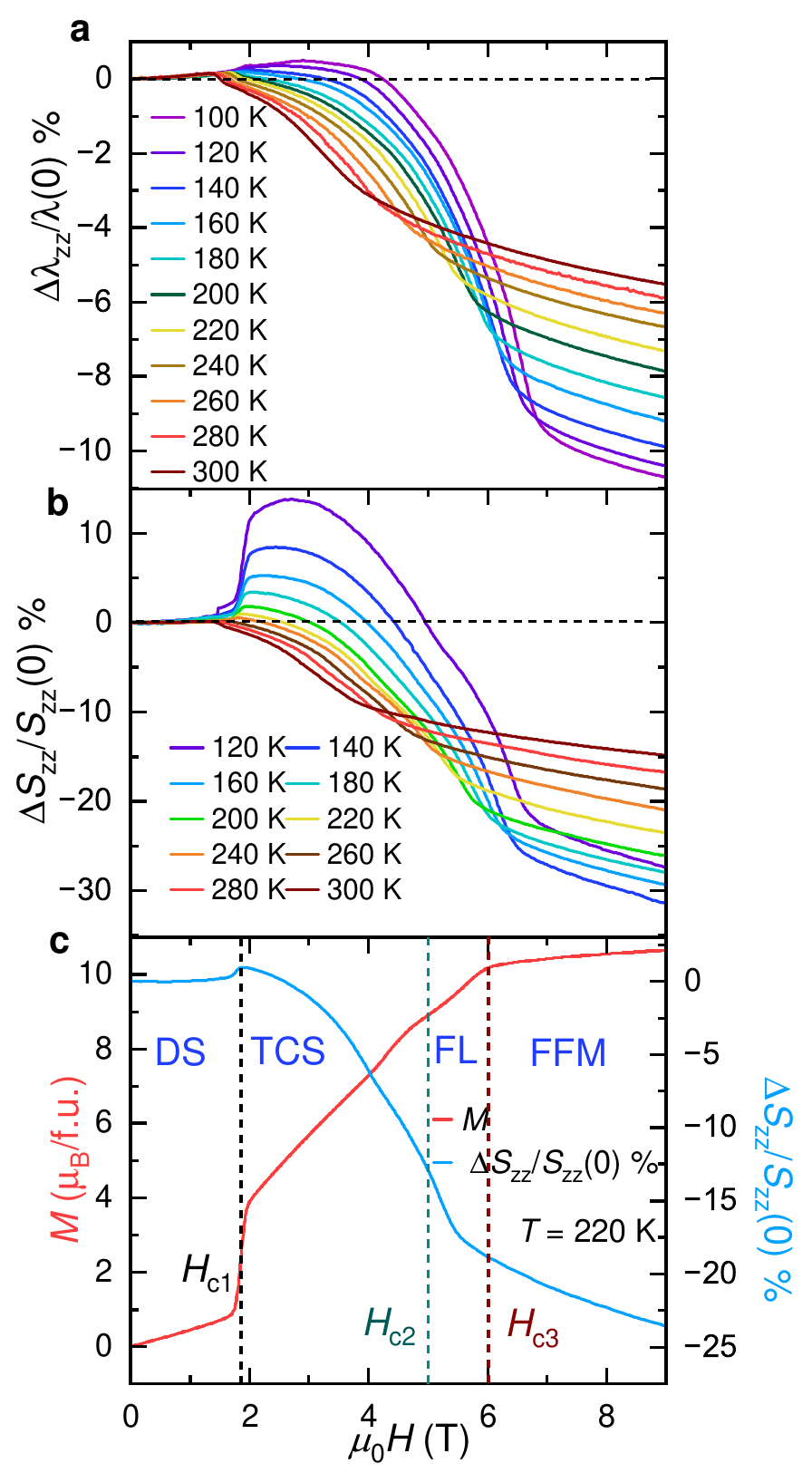}
	\caption{Magnetic field dependence of change in the longitudinal thermal resistivity and change in the Seebeck coefficient for various temperatures is shown in (a) and (b), respectively. (c) Comparison between the magnetization curve and change in Seebeck coefficient at \textit{T} = 220 K is shown. Both the curves go through change in slope as the external applied field is increased signifying the evolution of the magnetic structure: DS to TCS to FL to FFM in ScMn$_6$Sn$_6$.
 } \label{Fig3}
\end{figure} 
\par Further, we investigate the effect of magnetic field on the longitudinal thermal and thermoelectric properties of ScMn$_6$Sn$_6$. Figures~\Ref{Fig3}(a) and (b) show the magnetic field dependence of change in thermal resistivity ($\Delta \lambda_{zz}/ \lambda_{zz} (0)$) and change in $S_{zz}$ for $\mu_0H$= 0 $-$ 9 T and $T$ = 100 $-$ 300 K. $\Delta \lambda_{zz}/ \lambda_{zz} (0)$ manifests not only electrons, but also the charge neutral quasi-particle contributions such as phonon and magnon carrying heat.%suggests that electrons, phonons and magnons, all play a significant role as heat carriers at all measured temperatures.
The change in $\Delta \lambda_{zz}/ \lambda_{zz} (0)$ and $\Delta S_{zz}/S_{zz}(0)$ is almost negligible for $\mu_0H$ $<$ 2 T, followed by two slope changes at critical fields $H_{c1}$ and $H_{c2}$ as we increase the magnetic field beyond 2 T before it saturates at fields higher than $H_{c3}$. $H_{c1}$, $H_{c2}$ and $H_{c3}$ are the critical fields corresponding to the magnetic phase transitions from DS--TCS, TCS--FL and FL--FFM, respectively. The slope changes in $\Delta \lambda_{zz}/ \lambda_{zz} (0)$ and $\Delta S_{zz}/S_{zz}(0)$ occur at the magnetic phase transitions which indicates the strong coupling between thermal/thermoelectric properties and magnetic structures.
%The presence of slope changes at field values associated with the magnetic phase transitions leads us to conclude that there is a strong correlation between the thermal/thermoelectric properties and exotic magnetic structure of ScMn$_6$Sn$_6$. 
To show this relationship more clearly we compare between the magnetization and $\Delta S_{zz}/S_{zz}(0)$ curves at $T$ = 220 K [Figure~\Ref{Fig3}(c)]. The dashed lines help to see the one to one correspondence between the slope changes of the magnetization and $\Delta S_{zz}/S_{zz}(0)$ curves. In the past, reports of similar study on different magnetic materials with $\Delta S_{zz}/S_{zz}(0)$ showing identical trend has also been linked with the presence of non-trivial magnetic textures~\cite{Fe3Sn2,MnBi2Te4,MnGe-thermopower,YMn6Sn6_Nernst}. According to the preliminary studies of the relationship between the Seebeck coefficient and magnetism, there are two representative mechanisms that explain the change in the slope of $\Delta S_{zz}/S_{zz}(0)$:  %The slope change at the magnetic structure cross-overs is a result of two mechanisms: 
1) the alignment of the spins due to the external magnetic field reduces the spin scattering process which then affects the charge dynamics; 2) as stated earlier, $ S_{zz}$ can be explained in terms of average entropy of the charge carriers. So the large entropy change that occurs during any magnetic phase evolution contributes to the $\Delta S_{zz}/S_{zz}(0)$. 
\par On a closer inspection we see that for $T$ $\leq$ 200 K the percentage change of thermopower becomes positive within a certain field range in the TCS phase as $\mu_0H$ is increased above $H_{c1}$. The strength of the non-negative signal becomes stronger as the system is cooled below 200 K. In general, for a magnetic system $ S_{zz}$ should decrease with the magnetic field. The unconventional positive behavior of the thermopower with field has been previously reported in MnGe~\cite{MnGe-thermopower} system as well, where the authors relate the increase in $ S_{zz}$ to the enhanced magnetic fluctuations at the phase boundary resulting in the energy dependence 
of the transport lifetime. In the case of MnGe, a positive magnetoresistance (MR) was also reported alongside $ S_{zz}$. However, in ScMn$_6$Sn$_6$, MR shows almost negligible positive signal with the increase in $\mu_0H$ [see Figure S3 supplementary]. The response of MR and $\Delta S_{zz}/S_{zz}(0)$ to magnetic field is opposite within the TCS phase below $T$ = 200 K. This anticorrelation between MR and Seebeck thermopower was also seen in magnetic topological insulator MnBi$_2$Te$_4$~\cite{MnBi2Te4}. In the present case, we believe that there is some scattering phenomenon within the TCS phase that allows the carriers to gain high entropy values resulting in positive $\Delta S_{zz}/S_{zz}(0)$. Since the enhancement of the signal appears below 200 K, it further suggests that reduction in thermal fluctuations has a major role to play. Nevertheless, this scattering process is not detectable in the electrical and thermal resistivity. Further theoretical investigation is necessary to identify the exact mechanism behind the $\Delta S_{zz}/S_{zz}(0)$ vs. $\mu_0H$ trend.
\begin{figure*}[hbt]
	\includegraphics[width=\textwidth]{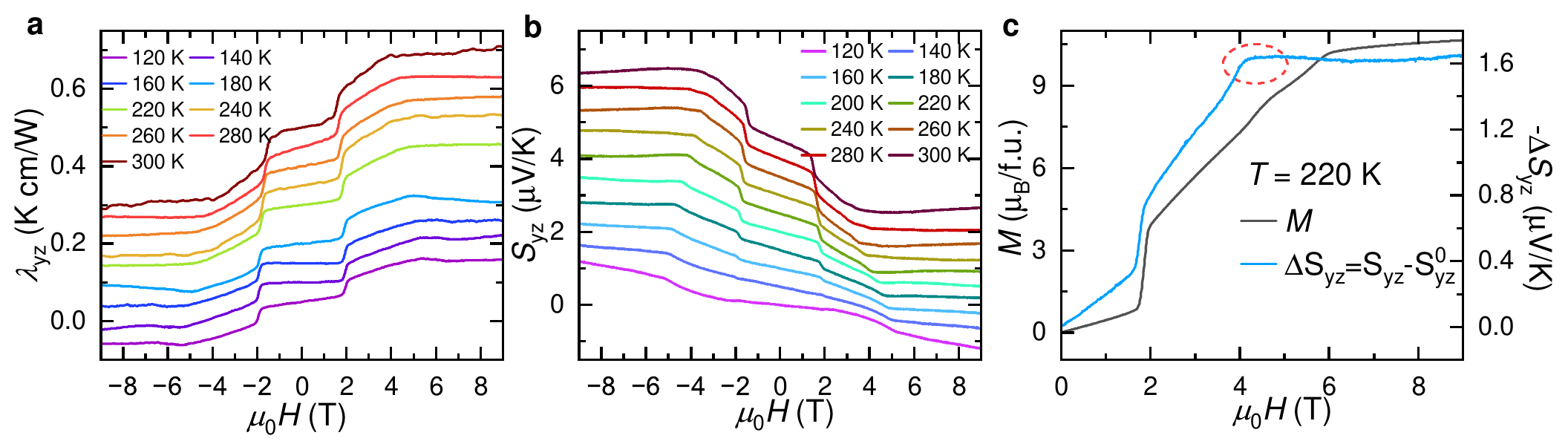}
	\caption{Magnetic field dependent thermal Hall resistivity ($\lambda_{yz}$) and Nernst coefficient($S_{yz}$) for various temperatures is shown in (a) and (b),respectively. The coefficient yz indicates that applied temperature gradient, $-$$\Delta T$ $||$ $z$-axis and the measured transverse $\Delta T$/thermoelectric voltage $||$ $y$-axis. (c) Comparison between the magnetization curve and residual Nernst coefficient at $T$ = 220 K is shown. Residual Nernst coefficient is obtained by subtracting the normal Nernst ($\propto H$) from the measured Nernst coefficient.
 } \label{Fig4}
\end{figure*} 
\par Next, we examine the field dependence of the transverse component of the thermal and thermoelectric signal. Magnetic field dependent thermal Hall resistivity ($\lambda_{yz}$) and Nernst coefficient ($S_{yz}$) for $\mu_0H$ = 9 to $-$9 T measured at several temperatures has been shown in Figures~\Ref{Fig4}(a) and (b), respectively. For clarity, each $\lambda_{yz}$ and $S_{yz}$ curves have been shifted with a certain constant value along the $\lambda_{yz}$/$S_{yz}$ axis. In Figure~\Ref{Fig4}(a) the observed $\lambda_{yz}$ vs. $\mu_0H$  follows the behavior of the field dependent magnetization. Similar to the $M$ vs. $\mu_0H$, $\lambda_{yz}$ vs. $\mu_0H$ clearly depicts the DS--TCS magnetic phase transition with a sharp jump at $H_{c1}$ at all measured temperatures, and saturates at $\mu_0H$ $\geq$ $H_{c3}$. The resemblance of the thermal Hall resistivity with that of magnetization curve distinctly indicates the presence of ATHE in ScMn$_6$Sn$_6$. As mentioned in the introduction, the observation of anomalous behavior correlates to the BC. Hence the presence of ATHE in ScMn$_6$Sn$_6$ implies the presence of large BC. ATHE has been previously reported in isostructural ferrimagnetic TbMn$_6$Sn$_6$ where its occurrence was related to the BC arising from the massive Dirac gaps~\cite{TbMn6Sn6-thermaltransport}. 
\par The field dependent Nernst thermopower [Figure~\Ref{Fig4}(b)], which is experimentally obtained as $S_{yz}$ = $E_y$/$\nabla_zT$, also shows identical features as the magnetization curve indicating the presence of ANE. The sign of $S_{yz}$ is in agreement with the reports of YMn$_6$Sn$_6$ and TbMn$_6$Sn$_6$ ~\cite{YMn6Sn6_Nernst,TbMn6Sn6-2ndthermaltransport}. To scrutinize the measured Nernst coefficient, we express it in terms of its constituents: $S_{yz}$ = $S_{yz}^0$ + $\Delta S_{yz}$, where $S_{yz}^0$ is the ordinary Nernst signal which is proportional to the applied magnetic field and $\Delta S_{yz}$ is the residual Nernst coefficient. Since, $S_{yz}^0$ $\propto H$, the ordinary contribution can be subtracted from the measured signal by performing the linear fit using the high magnetic field data. After reduction of the normal component from $S_{yz}$, we compare the residual signal at $T$ = 220 K with the magnetization. Figure~\Ref{Fig4}(c) shows $\Delta S_{yz}$ vs. $\mu_0H$ and $M$ vs. $\mu_0H$ at $T$ = 220 K. Although it seems that the $\Delta S_{yz}$ follows $M$, on a closer inspection it is distinct that there are some additional features in $\Delta S_{yz}$ compared to $M$ as indicated by the red-dashed circle in the Figure~\Ref{Fig4}(c). In general, any superimposed component on the transverse signal which doesn't follow the magnetization behavior is identified as a topological effect.%This extra attribute in $\Delta S_{yz}$ motivates us to investigate for the geometrical effects in $S_{yz}$.
\begin{figure}[hbt]
	\includegraphics[width= 0.34\textwidth]{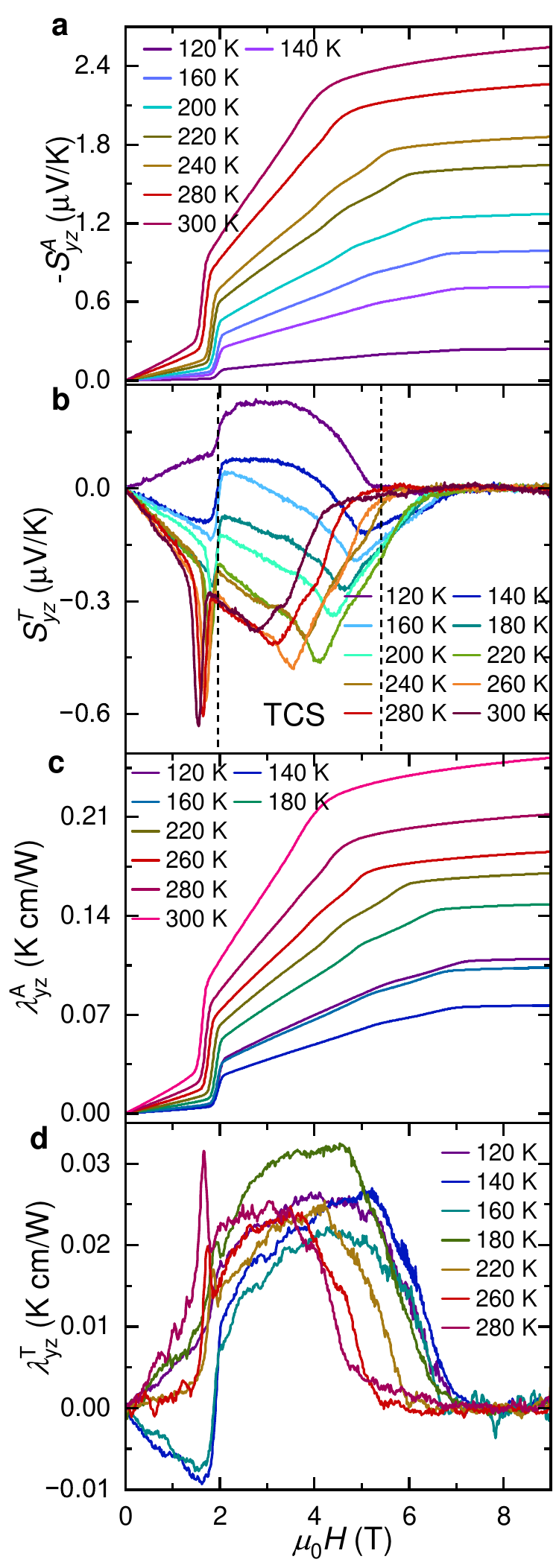}
	
	\caption{Anomalous and topological effects. (a) Anomalous Nernst response ($\propto M$) obtained from the measured Nernst signal is shown as a function of magnetic field for various temperatures. (b) Magnetic field dependence of the Nernst signal after subtracting the ordinary ($\propto H$) and anomalous response ($\propto M$). Field dependence of the anomalous (c) and topological (d) Hall contributions to the total thermal Hall signal at various temperatures.
 } \label{Fig5}
\end{figure} 

\begin{table}[hbt]
\begin{tabular}{|l|l|l|}
\hline
\textbf{Compound} & \textbf{$S_{yz}^A$}      & \textbf{Ref.} \\ \hline
Co$_2$MnGa        & $\sim$ 6 $\mu$VK$^{-1}$ & \cite{Co2MnGa}             \\ \hline
YMn$_6$Sn$_6$     & $\sim$ 2 $\mu$VK$^{-1}$ & \cite{YMn6Sn6_Nernst}             \\ \hline
TbMn$_6$Sn$_6$    & 2.2 $\mu$VK$^{-1}$      &\cite{TbMn6Sn6-2ndthermaltransport,TbMn6Sn6-thermaltransport}              \\ \hline
ScMn$_6$Sn$_6$    & 2.21 $\mu$VK$^{-1}$     & This work    \\ \hline
\end{tabular}
\caption{Comparison between the magnitude of the Nernst coefficient at $T$ = 300 K}\label{Tab1}
\end{table}
\par The Nernst thermopower can further be decomposed into three components: $S_{yz}$ = $S_{yz}^0$ + $S_{yz}^A$ + $S_{yz}^T$, where $S_{yz}^A$ is the anomalous contribution and $S_{yz}^T$ is the topological addition. With the assumption that $S_{yz}^A$ $\propto M$ and $S_{yz}^0$ $\propto H$, $S_{yz}^T$ can be extracted from the measured Nernst thermopower for each temperature. Figures~\Ref{Fig5}(a) and (b) report the educed $-S_{yz}^A$ and $S_{yz}^T$ signals with respect to field at various temperatures, respectively. As expected, the field dependent $-S_{yz}^A$ perfectly compares with the magnetization curve. 
%which reflects that the contribution from the Berry curvature alters around the magnetic phase transition.
The anomalous Nernst coefficient was estimated by extrapolating the high field data to zero-field value [Supplementary Figure S4(a)]. The magnitude of $S_{yz}^A$ at $T$ = 300 K is 2.21 $\mu$V/K which is very similar to the value reported for YMn$_6$Sn$_6$ and TbMn$_6$Sn$_6$ and comparable to the highest value reported for semimetals Co$_2$MnGa [Table~\Ref{Tab1}]. In contrast to YMn$_6$Sn$_6$~\cite{YMn6Sn6_Nernst}, the magnitude of $S_{yz}^A$ increases as $T$ is raised from 120 $-$ 300 K [Figure S4(b) supplementary], indicating a significant contribution from the BC at the Fermi level up to the highest measured temperature. 
%Similar temperature profile of $S_{yz}^A$ was observed in TbMn$_6$Sn$_6$~\cite{TbMn6Sn6-2ndthermaltransport,TbMn6Sn6-thermaltransport}. 
Figure~\Ref{Fig5}(b) confirms the occurrence of topological Nernst effect, $S_{yz}^T$, in ScMn$_6$Sn$_6$ at all measured temperatures. The magnitude of the response is maximum within the TCS magnetic phase of ScMn$_6$Sn$_6$. The extremes of the TCS phase is shown by the two black-dashed lines [Note: the lower and upper boundaries were identified using $T$ = 300 K and $T$ = 120 K magnetization curves, respectively]. The TCS phase in this system is also responsible for the origin of topological Hall effect as reported by Zhang $et$ $al$~\cite{ScMn6Sn6-APL}. The authors in the paper corroborate the theory put forward for YMn$_6$Sn$_6$~\cite{YMn6Sn6-Nirmal} which states that the non-collinear TCS phase gives rise to a dynamical non-zero spin chirality resulting in the deflection of conducting charge carriers and hence producing superimposed geometrical effects on the Hall component. 
As TNE and THE are both present in the TCS phase of ScMn$_6$Sn$_6$ we believe their emergence is correlated and they should follow the same theory. Nevertheless, TNE in the TCS phase is not revealed in YMn$_6$Sn$_6$~\cite{YMn6Sn6_Nernst}. Although, the two compounds,YMn$_6$Sn$_6$ and ScMn$_6$Sn$_6$, share similar magnetic structure and electrical transport properties, the absence of TNE in YMn$_6$Sn$_6$ suggests that there might be some subtle differences in the spin configuration of the TCS phase of the two materials which needs further exploration. The peak seen for $\mu_0H$ $<$ 2 T [Figure~\Ref{Fig5}(b)] is probably due to some mismatch in the critical fields responsible for DS to TCS magnetic cross-overs as seen by two completely different measurements, magnetization and thermoelectric transport [Fig. S5 supplementary]. Furthermore, below $T$ = 200 K, $S_{yz}^T$ starts to form a concave downward curvature, and eventually changes direction for $T$= 120 K. This $T$ = 200 K coincides with the temperature where $\Delta S_{zz}$ starts to show unconventional behavior [Figure~\Ref{Fig3}(b)], however, we are unable to establish a common phenomenon for this synchronicity. The change in sign of $S_{yz}^T$ at lower temperature was observed in Fe$_3$Sn$_2$ as well. 
\par Likewise, if we consider that similar to electrical Hall and Nernst thermopower, the total thermal Hall resistivity, $\lambda_{yz}$, can also be written as a sum of normal, anomalous and topological components, i.e. $\lambda_{yz}$ = $\lambda_{yz}^0$ + $\lambda_{yz}^A$ + $\lambda_{yz}^T$, then we can extract the values of $\lambda_{yz}^A$ and $\lambda_{yz}^T$ by following the procedure followed for $S_{yz}^A$ and $S_{yz}^T$. Here we assume that at high fields the phonon contribution to $\lambda_{yz}$ is linear in field ($\lambda_{yz}^0 \propto H$) and $\lambda_{yz}^A \propto M$. Figures~\Ref{Fig5}(c) and (d) show the field dependent obtained values for $\lambda_{yz}^A$ and $\lambda_{yz}^T$, respectively, at various temperatures. As expected $\lambda_{yz}^A$ behaves consistent with the magnetization. Surprisingly, a significant amount of signal for the topological thermal Hall effect is detected for $T$= 120 - 280 K [Figure~\Ref{Fig5} (d)]. The magnitude of $\lambda_{yz}^T$ becomes almost negligible at $T$ = 300 K [Supplemental Figure S6]. Although there has been theoretical predictions for the existence of TTHE in kagome~\cite{TTHE-kagomeAFMs} and frustrated antiferromagnets~\cite{TTHE-frustratedAFMs}, identifying it through experiments has been a great challenge. Experimentally, TTHE has been previously reported in a skyrmion lattice system GaV$_4$Se$_8$~\cite{GaV4Se8}, a magnetic insulator, below 15 K. It is theoretically predicted that the motion of the non-collinear spin textures and magnons are coupled in the presence of fictitious electromechanical potential~\cite{magnetictextureinducedTHE,magnondeflectviaskyrmion}. Since maxima of $\lambda_{yz}^T$ is also limited to the TCS phase, this makes us suspect that in ScMn$_6$Sn$_6$, the non-zero spin chirality in the TCS phase also deflects the quasi particle like magnons  which carry thermal currents along with electrons, resulting in $\lambda_{yz}^T$. This is the first time such effect is seen in a non-skyrmion system experimentally up to room temperature. The exact mechanism responsible for generating TTHE in ScMn$_6$Sn$_6$ is to be explored in future.
\par
\begin{figure}[hbt]
	\includegraphics[width= 8.6 cm]{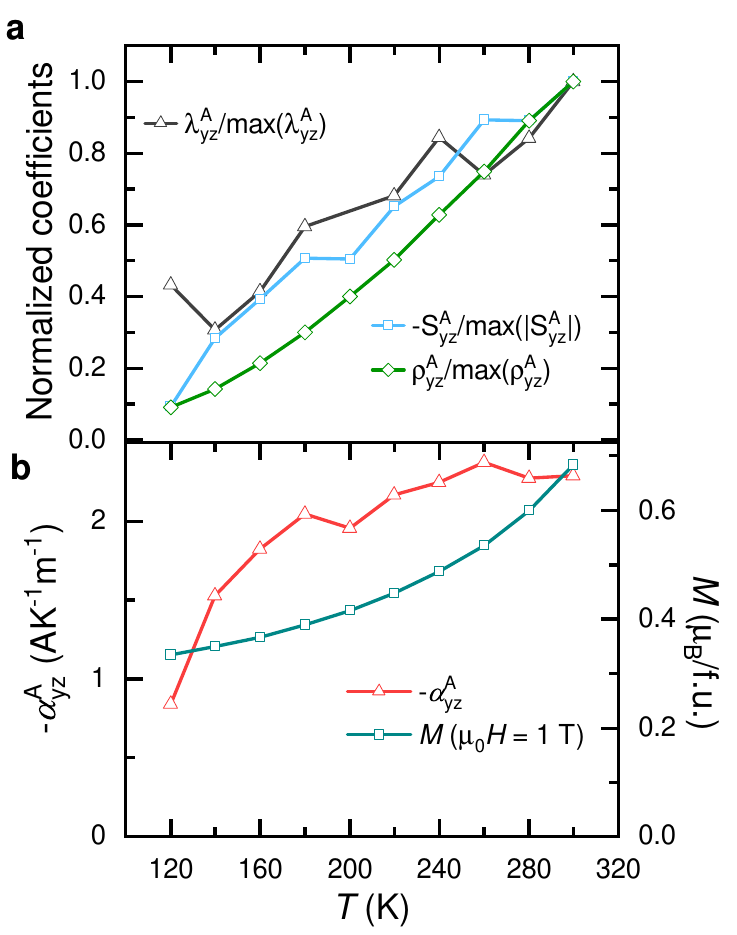}
	
	\caption{(a) Temperature evolution of the various normalized coefficients: $\lambda_{yz}^A$,$-S_{yz}^A$, and $\rho_{yz}^A$ at 7 T. (b) Thermal dependence of  anomalous thermoelectric linear response tensor ($\alpha_{yz}^A$) and magnetization at 1 T.
 } \label{Fig6}
\end{figure}

 To get a deeper insight about the origin of the anomalous behavior in ScMn$_6$Sn$_6$ we compare the temperature evolution of coefficients obtained from three independent measurements: electrical, thermal, and thermoelectric, at 7 T, regime of saturated magnetic state. Figure~\Ref{Fig6}(a) shows the temperature progression of  $\lambda_{yz}^A$,$-S_{yz}^A$, and $\rho_{yz}^A$ at 7 T. Each of these quantities have been normalized to itself to see the trend more clearly. Although these transverse coefficients, $\lambda_{yz}^A$, $-S_{yz}^A$, and $\rho_{yz}^A$, have been acquired separately from thermoelectric, thermal and electrical measurements, respectively, their response to the increasing temperature remains the same. The magnitude of all three increase with temperature values from $T$ = 120 K to $T$ = 300 K, this similarity between the coefficients indicates that these effects have a common origin, the BC of ScMn$_6$Sn$_6$.   
\par Lastly, we define transverse component of the thermoelectric tensor, $\alpha_{yz}$, which allows us to correlate the electrical and thermoelectric transport properties. For a thermoelectric measurements, the electrical current flowing through the material is zero. In such a situation  $\alpha_{yz}$ can be defined as~\cite{forthermoelectrictensor,Co2MnGa,signofNenst}:
\begin{equation}\label{TE}
   \alpha_{yz} = (S_{yz}\sigma_{zz} + S_{zz}\sigma_{yz})
  \end{equation}
In terms of electrical resistivity the above equation can be re-written as:
\begin{equation}\label{TE2}
   \alpha_{yz} = \dfrac{S_{yz}\rho_{zz} - S_{zz}\rho_{yz} }{\rho_{zz}\rho_{yy} + \rho_{yz}^2}
  \end{equation}
such that $\alpha_{yz}^A$ = $\dfrac{S_{yz}^A\rho_{zz} - S_{zz}\rho_{yz}^A }{\rho_{zz}\rho_{yy} + \rho_{yz}^2}$.
Figure~\Ref{Fig6}(b) displays the temperature profile of the anomalous coefficient of the thermoelectric tensor, $-\alpha_{yz}^A$ and compares it with the magnetization ($M$) at $\mu_0 H$ = 1 T. $\alpha_{yz}^A$ is obtained by extrapolating the slope of high field data to zero field value [Supplementary Fig. S7]. For a conventional ferromagnetic system, it is required that $\alpha_{yz}^A$ scales with the temperature dependence of the magnetization~\cite{thermoelectricandMag,Co2MnGa}. In the present case, $\alpha_{yz}^A$ and $M$ don't scale linearly, however, similar to $M$ the magnitude of $\alpha_{yz}^A$ increases with the temperature. Moreover, $\alpha_{yz}^A$ vs. $T$ of ScMn$_6$Sn$_6$ doesn't follow the behavior of $\alpha_{yz}^A$ vs. $T$ of YMn$_6$Sn$_6$~\cite{YMn6Sn6_Nernst}. In YMn$_6$Sn$_6$ the magnitude of $\alpha_{yz}^A$ lowers with higher temperature. We don't have a clear understanding about the differences in  the thermal dependence of $\alpha_{yz}^A$ in these isostructural compounds.
%\par The electrical Hall measurements identified ScMn$_6$Sn$_6$ as identical to YMn$_6$Sn$_6$~\cite{ScMn6Sn6-APL}, while the transverse thermal/thermoelectric measurements shed light into some distinctness between these compounds.The two most important differences which have been observed through thermoelectric transport are: 1) evidence of additional scattering phenomenon below $T$ = 200 K in the TCS phase [Figure~\Ref{Fig3}(b)] and 2) observation of TNE in ScMn$_6$Sn$_6$.This directs to conclude that the thermal/thermoelectric measurements due to their sensitivity to the scattering and chirality of heat carriers is a powerful tool to uncover the physical properties. These techniques can not only be used for the insulators but also for metallic samples, like in the present case, to reveal information which is undetectable by the electrical measurements.
\section{Conclusion}\label{Conclusion}
 In conclusion, we explored thermal and thermoelectric properties of ScMn$_6$Sn$_6$. It displays thermal Hall effect and Nernst thermopower with a magnitude of 2.21 $\mu$V/K anomalous Nernst signal at $T$ = 300 K. Field dependent study  on ScMn$_6$Sn$_6$ demonstrates presence of an unconventional Seebeck thermopower behavior below $T$ = 200 K, which is not usually observed in magnetic materials. Most importantly we revealed that this material not just exhibits topological Hall effect~\cite{ScMn6Sn6-APL}, but also topological Nernst and topological thermal Hall effects in the transverse conical spiral magnetic phase. Both these effects are extremely difficult phenomena  to discern. This makes ScMn$_6$Sn$_6$ a unique and rare example of skyrmion free system which manifests all three topological effects: electrical, thermal and thermoelectric. Further investigations are necessary to unravel the understanding about the presence of topological effects in ScMn$_6$Sn$_6$. This study opens up possibility to explore these effects in new materials with chiral spin fluctuations via thermal and thermoelectric measurement techniques.

\section*{Acknowledgment}
 RPM, SM and DGM acknowledge the support from AFOSR MURI (Novel Light-Matter Interactions in Topologically Non-Trivial Weyl Semimetal Structures and Systems), grant FA9550-20-1-0322. WRM, SD, RX, and TM acknowledge support from the Gordon and Betty Moore Foundation's EPiQS Initiative, Grant GBMF9069 to DGM. HZ acknowledges support from U.S. Department of Energy, Office of Science, Basic Energy Sciences, Materials Sciences and Engineering Division.

 \section*{References}

   % \bibliography{Refs2}

\begin{thebibliography}{44}%
   	\makeatletter
   	\providecommand \@ifxundefined [1]{%
   		\@ifx{#1\undefined}
   	}%
   	\providecommand \@ifnum [1]{%
   		\ifnum #1\expandafter \@firstoftwo
   		\else \expandafter \@secondoftwo
   		\fi
   	}%
   	\providecommand \@ifx [1]{%
   		\ifx #1\expandafter \@firstoftwo
   		\else \expandafter \@secondoftwo
   		\fi
   	}%
   	\providecommand \natexlab [1]{#1}%
   	\providecommand \enquote  [1]{``#1''}%
   	\providecommand \bibnamefont  [1]{#1}%
   	\providecommand \bibfnamefont [1]{#1}%
   	\providecommand \citenamefont [1]{#1}%
   	\providecommand \href@noop [0]{\@secondoftwo}%
   	\providecommand \href [0]{\begingroup \@sanitize@url \@href}%
   	\providecommand \@href[1]{\@@startlink{#1}\@@href}%
   	\providecommand \@@href[1]{\endgroup#1\@@endlink}%
   	\providecommand \@sanitize@url [0]{\catcode `\\12\catcode `\$12\catcode
   		`\&12\catcode `\#12\catcode `\^12\catcode `\_12\catcode `\%12\relax}%
   	\providecommand \@@startlink[1]{}%
   	\providecommand \@@endlink[0]{}%
   	\providecommand \url  [0]{\begingroup\@sanitize@url \@url }%
   	\providecommand \@url [1]{\endgroup\@href {#1}{\urlprefix }}%
   	\providecommand \urlprefix  [0]{URL }%
   	\providecommand \Eprint [0]{\href }%
   	\providecommand \doibase [0]{https://doi.org/}%
   	\providecommand \selectlanguage [0]{\@gobble}%
   	\providecommand \bibinfo  [0]{\@secondoftwo}%
   	\providecommand \bibfield  [0]{\@secondoftwo}%
   	\providecommand \translation [1]{[#1]}%
   	\providecommand \BibitemOpen [0]{}%
   	\providecommand \bibitemStop [0]{}%
   	\providecommand \bibitemNoStop [0]{.\EOS\space}%
   	\providecommand \EOS [0]{\spacefactor3000\relax}%
   	\providecommand \BibitemShut  [1]{\csname bibitem#1\endcsname}%
   	\let\auto@bib@innerbib\@empty
   	%</preamble>
   	\bibitem [{\citenamefont {Ikhlas}\ \emph {et~al.}(2017)\citenamefont {Ikhlas},
   		\citenamefont {Tomita}, \citenamefont {Koretsune}, \citenamefont {Suzuki},
   		\citenamefont {Nishio-Hamane}, \citenamefont {Arita}, \citenamefont {Otani},\
   		and\ \citenamefont {Nakatsuji}}]{largeANE-AFM}%
   	\BibitemOpen
   	\bibfield  {author} {\bibinfo {author} {\bibfnamefont {M.}~\bibnamefont
   			{Ikhlas}}, \bibinfo {author} {\bibfnamefont {T.}~\bibnamefont {Tomita}},
   		\bibinfo {author} {\bibfnamefont {T.}~\bibnamefont {Koretsune}}, \bibinfo
   		{author} {\bibfnamefont {M.-T.}\ \bibnamefont {Suzuki}}, \bibinfo {author}
   		{\bibfnamefont {D.}~\bibnamefont {Nishio-Hamane}}, \bibinfo {author}
   		{\bibfnamefont {R.}~\bibnamefont {Arita}}, \bibinfo {author} {\bibfnamefont
   			{Y.}~\bibnamefont {Otani}},\ and\ \bibinfo {author} {\bibfnamefont
   			{S.}~\bibnamefont {Nakatsuji}},\ }\href@noop {} {\bibfield  {journal}
   		{\bibinfo  {journal} {Nature Physics}\ }\textbf {\bibinfo {volume} {13}},\
   		\bibinfo {pages} {1085} (\bibinfo {year} {2017})}\BibitemShut {NoStop}%
   	\bibitem [{\citenamefont {Kolincio}\ \emph {et~al.}(2021)\citenamefont
   		{Kolincio}, \citenamefont {Hirschberger}, \citenamefont {Masell},
   		\citenamefont {Gao}, \citenamefont {Kikkawa}, \citenamefont {Taguchi},
   		\citenamefont {hisa Arima}, \citenamefont {Nagaosa},\ and\ \citenamefont
   		{Tokura}}]{TNE-Nd3Ru4Al12}%
   	\BibitemOpen
   	\bibfield  {author} {\bibinfo {author} {\bibfnamefont {K.~K.}\ \bibnamefont
   			{Kolincio}}, \bibinfo {author} {\bibfnamefont {M.}~\bibnamefont
   			{Hirschberger}}, \bibinfo {author} {\bibfnamefont {J.}~\bibnamefont
   			{Masell}}, \bibinfo {author} {\bibfnamefont {S.}~\bibnamefont {Gao}},
   		\bibinfo {author} {\bibfnamefont {A.}~\bibnamefont {Kikkawa}}, \bibinfo
   		{author} {\bibfnamefont {Y.}~\bibnamefont {Taguchi}}, \bibinfo {author}
   		{\bibfnamefont {T.}~\bibnamefont {hisa Arima}}, \bibinfo {author}
   		{\bibfnamefont {N.}~\bibnamefont {Nagaosa}},\ and\ \bibinfo {author}
   		{\bibfnamefont {Y.}~\bibnamefont {Tokura}},\ }\href@noop {} {\bibfield
   		{journal} {\bibinfo  {journal} {Proceedings of the National Academy of
   				Sciences}\ }\textbf {\bibinfo {volume} {118}},\ \bibinfo {pages}
   		{e2023588118} (\bibinfo {year} {2021})}\BibitemShut {NoStop}%
   	\bibitem [{\citenamefont {Wuttke}\ \emph {et~al.}(2019)\citenamefont {Wuttke},
   		\citenamefont {Caglieris}, \citenamefont {Sykora}, \citenamefont
   		{Scaravaggi}, \citenamefont {Wolter}, \citenamefont {Manna}, \citenamefont
   		{S\"uss}, \citenamefont {Shekhar}, \citenamefont {Felser}, \citenamefont
   		{B\"uchner},\ and\ \citenamefont {Hess}}]{Mn3Ge_ANE}%
   	\BibitemOpen
   	\bibfield  {author} {\bibinfo {author} {\bibfnamefont {C.}~\bibnamefont
   			{Wuttke}}, \bibinfo {author} {\bibfnamefont {F.}~\bibnamefont {Caglieris}},
   		\bibinfo {author} {\bibfnamefont {S.}~\bibnamefont {Sykora}}, \bibinfo
   		{author} {\bibfnamefont {F.}~\bibnamefont {Scaravaggi}}, \bibinfo {author}
   		{\bibfnamefont {A.~U.~B.}\ \bibnamefont {Wolter}}, \bibinfo {author}
   		{\bibfnamefont {K.}~\bibnamefont {Manna}}, \bibinfo {author} {\bibfnamefont
   			{V.}~\bibnamefont {S\"uss}}, \bibinfo {author} {\bibfnamefont
   			{C.}~\bibnamefont {Shekhar}}, \bibinfo {author} {\bibfnamefont
   			{C.}~\bibnamefont {Felser}}, \bibinfo {author} {\bibfnamefont
   			{B.}~\bibnamefont {B\"uchner}},\ and\ \bibinfo {author} {\bibfnamefont
   			{C.}~\bibnamefont {Hess}},\ }\href@noop {} {\bibfield  {journal} {\bibinfo
   			{journal} {Phys. Rev. B}\ }\textbf {\bibinfo {volume} {100}},\ \bibinfo
   		{pages} {085111} (\bibinfo {year} {2019})}\BibitemShut {NoStop}%
   	\bibitem [{\citenamefont {Guo}\ \emph {et~al.}(2022)\citenamefont {Guo},
   		\citenamefont {Xu}, \citenamefont {Cheng}, \citenamefont {Zhou},\ and\
   		\citenamefont {Chen}}]{usedforintro}%
   	\BibitemOpen
   	\bibfield  {author} {\bibinfo {author} {\bibfnamefont {S.}~\bibnamefont
   			{Guo}}, \bibinfo {author} {\bibfnamefont {Y.}~\bibnamefont {Xu}}, \bibinfo
   		{author} {\bibfnamefont {R.}~\bibnamefont {Cheng}}, \bibinfo {author}
   		{\bibfnamefont {J.}~\bibnamefont {Zhou}},\ and\ \bibinfo {author}
   		{\bibfnamefont {X.}~\bibnamefont {Chen}},\ }\href@noop {} {\bibfield
   		{journal} {\bibinfo  {journal} {The Innovation}\ }\textbf {\bibinfo {volume}
   			{3}},\ \bibinfo {pages} {100290} (\bibinfo {year} {2022})}\BibitemShut
   	{NoStop}%
   	\bibitem [{\citenamefont {Boulanger}\ \emph {et~al.}(2020)\citenamefont
   		{Boulanger}, \citenamefont {Grissonnanche}, \citenamefont {Badoux},
   		\citenamefont {Allaire}, \citenamefont {Lefran{\c{c}}ois}, \citenamefont
   		{Legros}, \citenamefont {Gourgout}, \citenamefont {Dion}, \citenamefont
   		{Wang}, \citenamefont {Chen} \emph {et~al.}}]{thermal-hall-insulators}%
   	\BibitemOpen
   	\bibfield  {author} {\bibinfo {author} {\bibfnamefont {M.-E.}\ \bibnamefont
   			{Boulanger}}, \bibinfo {author} {\bibfnamefont {G.}~\bibnamefont
   			{Grissonnanche}}, \bibinfo {author} {\bibfnamefont {S.}~\bibnamefont
   			{Badoux}}, \bibinfo {author} {\bibfnamefont {A.}~\bibnamefont {Allaire}},
   		\bibinfo {author} {\bibfnamefont {{\'E}.}~\bibnamefont {Lefran{\c{c}}ois}},
   		\bibinfo {author} {\bibfnamefont {A.}~\bibnamefont {Legros}}, \bibinfo
   		{author} {\bibfnamefont {A.}~\bibnamefont {Gourgout}}, \bibinfo {author}
   		{\bibfnamefont {M.}~\bibnamefont {Dion}}, \bibinfo {author} {\bibfnamefont
   			{C.}~\bibnamefont {Wang}}, \bibinfo {author} {\bibfnamefont {X.}~\bibnamefont
   			{Chen}}, \emph {et~al.},\ }\href@noop {} {\bibfield  {journal} {\bibinfo
   			{journal} {Nature communications}\ }\textbf {\bibinfo {volume} {11}},\
   		\bibinfo {pages} {5325} (\bibinfo {year} {2020})}\BibitemShut {NoStop}%
   	\bibitem [{\citenamefont {Ideue}\ \emph {et~al.}(2017)\citenamefont {Ideue},
   		\citenamefont {Kurumaji}, \citenamefont {Ishiwata},\ and\ \citenamefont
   		{Tokura}}]{thermal-hall-insulators2}%
   	\BibitemOpen
   	\bibfield  {author} {\bibinfo {author} {\bibfnamefont {T.}~\bibnamefont
   			{Ideue}}, \bibinfo {author} {\bibfnamefont {T.}~\bibnamefont {Kurumaji}},
   		\bibinfo {author} {\bibfnamefont {S.}~\bibnamefont {Ishiwata}},\ and\
   		\bibinfo {author} {\bibfnamefont {Y.}~\bibnamefont {Tokura}},\ }\href@noop {}
   	{\bibfield  {journal} {\bibinfo  {journal} {Nature materials}\ }\textbf
   		{\bibinfo {volume} {16}},\ \bibinfo {pages} {797} (\bibinfo {year}
   		{2017})}\BibitemShut {NoStop}%
   	\bibitem [{\citenamefont {Saito}\ \emph {et~al.}(2019)\citenamefont {Saito},
   		\citenamefont {Misaki}, \citenamefont {Ishizuka},\ and\ \citenamefont
   		{Nagaosa}}]{thermal-hall-insulators3}%
   	\BibitemOpen
   	\bibfield  {author} {\bibinfo {author} {\bibfnamefont {T.}~\bibnamefont
   			{Saito}}, \bibinfo {author} {\bibfnamefont {K.}~\bibnamefont {Misaki}},
   		\bibinfo {author} {\bibfnamefont {H.}~\bibnamefont {Ishizuka}},\ and\
   		\bibinfo {author} {\bibfnamefont {N.}~\bibnamefont {Nagaosa}},\ }\href@noop
   	{} {\bibfield  {journal} {\bibinfo  {journal} {Phys. Rev. Lett.}\ }\textbf
   		{\bibinfo {volume} {123}},\ \bibinfo {pages} {255901} (\bibinfo {year}
   		{2019})}\BibitemShut {NoStop}%
   	\bibitem [{\citenamefont {Zhang}\ \emph {et~al.}(2021)\citenamefont {Zhang},
   		\citenamefont {Xu},\ and\ \citenamefont {Ke}}]{Fe3Sn2}%
   	\BibitemOpen
   	\bibfield  {author} {\bibinfo {author} {\bibfnamefont {H.}~\bibnamefont
   			{Zhang}}, \bibinfo {author} {\bibfnamefont {C.~Q.}\ \bibnamefont {Xu}},\ and\
   		\bibinfo {author} {\bibfnamefont {X.}~\bibnamefont {Ke}},\ }\href@noop {}
   	{\bibfield  {journal} {\bibinfo  {journal} {Phys. Rev. B}\ }\textbf {\bibinfo
   			{volume} {103}},\ \bibinfo {pages} {L201101} (\bibinfo {year}
   		{2021})}\BibitemShut {NoStop}%
   	\bibitem [{\citenamefont {Nagaosa}\ \emph {et~al.}(2010)\citenamefont
   		{Nagaosa}, \citenamefont {Sinova}, \citenamefont {Onoda}, \citenamefont
   		{MacDonald},\ and\ \citenamefont {Ong}}]{AHE-formula-MacDonald}%
   	\BibitemOpen
   	\bibfield  {author} {\bibinfo {author} {\bibfnamefont {N.}~\bibnamefont
   			{Nagaosa}}, \bibinfo {author} {\bibfnamefont {J.}~\bibnamefont {Sinova}},
   		\bibinfo {author} {\bibfnamefont {S.}~\bibnamefont {Onoda}}, \bibinfo
   		{author} {\bibfnamefont {A.~H.}\ \bibnamefont {MacDonald}},\ and\ \bibinfo
   		{author} {\bibfnamefont {N.~P.}\ \bibnamefont {Ong}},\ }\href@noop {}
   	{\bibfield  {journal} {\bibinfo  {journal} {Reviews of modern physics}\
   		}\textbf {\bibinfo {volume} {82}},\ \bibinfo {pages} {1539} (\bibinfo {year}
   		{2010})}\BibitemShut {NoStop}%
   	\bibitem [{\citenamefont {Asaba}\ \emph {et~al.}(2021)\citenamefont {Asaba},
   		\citenamefont {Ivanov}, \citenamefont {Thomas}, \citenamefont {Savrasov},
   		\citenamefont {Thompson}, \citenamefont {Bauer},\ and\ \citenamefont
   		{Ronning}}]{ANE-formula-UCo0.8Ru0.2Al}%
   	\BibitemOpen
   	\bibfield  {author} {\bibinfo {author} {\bibfnamefont {T.}~\bibnamefont
   			{Asaba}}, \bibinfo {author} {\bibfnamefont {V.}~\bibnamefont {Ivanov}},
   		\bibinfo {author} {\bibfnamefont {S.}~\bibnamefont {Thomas}}, \bibinfo
   		{author} {\bibfnamefont {S.}~\bibnamefont {Savrasov}}, \bibinfo {author}
   		{\bibfnamefont {J.}~\bibnamefont {Thompson}}, \bibinfo {author}
   		{\bibfnamefont {E.}~\bibnamefont {Bauer}},\ and\ \bibinfo {author}
   		{\bibfnamefont {F.}~\bibnamefont {Ronning}},\ }\href@noop {} {\bibfield
   		{journal} {\bibinfo  {journal} {Science Advances}\ }\textbf {\bibinfo
   			{volume} {7}},\ \bibinfo {pages} {eabf1467} (\bibinfo {year}
   		{2021})}\BibitemShut {NoStop}%
   	\bibitem [{\citenamefont {Guin}\ \emph {et~al.}(2019)\citenamefont {Guin},
   		\citenamefont {Manna}, \citenamefont {Noky}, \citenamefont {Watzman},
   		\citenamefont {Fu}, \citenamefont {Kumar}, \citenamefont {Schnelle},
   		\citenamefont {Shekhar}, \citenamefont {Sun}, \citenamefont {Gooth} \emph
   		{et~al.}}]{Co2MnGa}%
   	\BibitemOpen
   	\bibfield  {author} {\bibinfo {author} {\bibfnamefont {S.~N.}\ \bibnamefont
   			{Guin}}, \bibinfo {author} {\bibfnamefont {K.}~\bibnamefont {Manna}},
   		\bibinfo {author} {\bibfnamefont {J.}~\bibnamefont {Noky}}, \bibinfo {author}
   		{\bibfnamefont {S.~J.}\ \bibnamefont {Watzman}}, \bibinfo {author}
   		{\bibfnamefont {C.}~\bibnamefont {Fu}}, \bibinfo {author} {\bibfnamefont
   			{N.}~\bibnamefont {Kumar}}, \bibinfo {author} {\bibfnamefont
   			{W.}~\bibnamefont {Schnelle}}, \bibinfo {author} {\bibfnamefont
   			{C.}~\bibnamefont {Shekhar}}, \bibinfo {author} {\bibfnamefont
   			{Y.}~\bibnamefont {Sun}}, \bibinfo {author} {\bibfnamefont {J.}~\bibnamefont
   			{Gooth}}, \emph {et~al.},\ }\href@noop {} {\bibfield  {journal} {\bibinfo
   			{journal} {NPG Asia Materials}\ }\textbf {\bibinfo {volume} {11}},\ \bibinfo
   		{pages} {16} (\bibinfo {year} {2019})}\BibitemShut {NoStop}%
   	\bibitem [{\citenamefont {Liang}\ \emph {et~al.}(2017)\citenamefont {Liang},
   		\citenamefont {Lin}, \citenamefont {Gibson}, \citenamefont {Gao},
   		\citenamefont {Hirschberger}, \citenamefont {Liu}, \citenamefont {Cava},\
   		and\ \citenamefont {Ong}}]{signofNenst}%
   	\BibitemOpen
   	\bibfield  {author} {\bibinfo {author} {\bibfnamefont {T.}~\bibnamefont
   			{Liang}}, \bibinfo {author} {\bibfnamefont {J.}~\bibnamefont {Lin}}, \bibinfo
   		{author} {\bibfnamefont {Q.}~\bibnamefont {Gibson}}, \bibinfo {author}
   		{\bibfnamefont {T.}~\bibnamefont {Gao}}, \bibinfo {author} {\bibfnamefont
   			{M.}~\bibnamefont {Hirschberger}}, \bibinfo {author} {\bibfnamefont
   			{M.}~\bibnamefont {Liu}}, \bibinfo {author} {\bibfnamefont {R.~J.}\
   			\bibnamefont {Cava}},\ and\ \bibinfo {author} {\bibfnamefont {N.~P.}\
   			\bibnamefont {Ong}},\ }\href@noop {} {\bibfield  {journal} {\bibinfo
   			{journal} {Phys. Rev. Lett.}\ }\textbf {\bibinfo {volume} {118}},\ \bibinfo
   		{pages} {136601} (\bibinfo {year} {2017})}\BibitemShut {NoStop}%
   	\bibitem [{\citenamefont {Neubauer}\ \emph {et~al.}(2009)\citenamefont
   		{Neubauer}, \citenamefont {Pfleiderer}, \citenamefont {Binz}, \citenamefont
   		{Rosch}, \citenamefont {Ritz}, \citenamefont {Niklowitz},\ and\ \citenamefont
   		{B\"oni}}]{THE-MnSi}%
   	\BibitemOpen
   	\bibfield  {author} {\bibinfo {author} {\bibfnamefont {A.}~\bibnamefont
   			{Neubauer}}, \bibinfo {author} {\bibfnamefont {C.}~\bibnamefont
   			{Pfleiderer}}, \bibinfo {author} {\bibfnamefont {B.}~\bibnamefont {Binz}},
   		\bibinfo {author} {\bibfnamefont {A.}~\bibnamefont {Rosch}}, \bibinfo
   		{author} {\bibfnamefont {R.}~\bibnamefont {Ritz}}, \bibinfo {author}
   		{\bibfnamefont {P.~G.}\ \bibnamefont {Niklowitz}},\ and\ \bibinfo {author}
   		{\bibfnamefont {P.}~\bibnamefont {B\"oni}},\ }\href@noop {} {\bibfield
   		{journal} {\bibinfo  {journal} {Phys. Rev. Lett.}\ }\textbf {\bibinfo
   			{volume} {102}},\ \bibinfo {pages} {186602} (\bibinfo {year}
   		{2009})}\BibitemShut {NoStop}%
   	\bibitem [{\citenamefont {Akazawa}\ \emph {et~al.}(2022)\citenamefont
   		{Akazawa}, \citenamefont {Lee}, \citenamefont {Takeda}, \citenamefont
   		{Fujima}, \citenamefont {Tokunaga}, \citenamefont {Arima}, \citenamefont
   		{Han},\ and\ \citenamefont {Yamashita}}]{GaV4Se8}%
   	\BibitemOpen
   	\bibfield  {author} {\bibinfo {author} {\bibfnamefont {M.}~\bibnamefont
   			{Akazawa}}, \bibinfo {author} {\bibfnamefont {H.-Y.}\ \bibnamefont {Lee}},
   		\bibinfo {author} {\bibfnamefont {H.}~\bibnamefont {Takeda}}, \bibinfo
   		{author} {\bibfnamefont {Y.}~\bibnamefont {Fujima}}, \bibinfo {author}
   		{\bibfnamefont {Y.}~\bibnamefont {Tokunaga}}, \bibinfo {author}
   		{\bibfnamefont {T.-h.}\ \bibnamefont {Arima}}, \bibinfo {author}
   		{\bibfnamefont {J.~H.}\ \bibnamefont {Han}},\ and\ \bibinfo {author}
   		{\bibfnamefont {M.}~\bibnamefont {Yamashita}},\ }\href@noop {} {\bibfield
   		{journal} {\bibinfo  {journal} {Phys. Rev. Res.}\ }\textbf {\bibinfo {volume}
   			{4}},\ \bibinfo {pages} {043085} (\bibinfo {year} {2022})}\BibitemShut
   	{NoStop}%
   	\bibitem [{\citenamefont {Nakamura}\ \emph {et~al.}()\citenamefont {Nakamura},
   		\citenamefont {Morikawa}, \citenamefont {Yu}, \citenamefont {Kagawa},
   		\citenamefont {Arima}, \citenamefont {Tokura},\ and\ \citenamefont
   		{Kawasaki}}]{THE_manganites}%
   	\BibitemOpen
   	\bibfield  {author} {\bibinfo {author} {\bibfnamefont {M.}~\bibnamefont
   			{Nakamura}}, \bibinfo {author} {\bibfnamefont {D.}~\bibnamefont {Morikawa}},
   		\bibinfo {author} {\bibfnamefont {X.}~\bibnamefont {Yu}}, \bibinfo {author}
   		{\bibfnamefont {F.}~\bibnamefont {Kagawa}}, \bibinfo {author} {\bibfnamefont
   			{T.-h.}\ \bibnamefont {Arima}}, \bibinfo {author} {\bibfnamefont
   			{Y.}~\bibnamefont {Tokura}},\ and\ \bibinfo {author} {\bibfnamefont
   			{M.}~\bibnamefont {Kawasaki}},\ }\href@noop {} {\bibfield  {journal}
   		{\bibinfo  {journal} {Journal of the Physical Society of Japan}\ }\textbf
   		{\bibinfo {volume} {87}},\ \bibinfo {pages} {074704}}\BibitemShut {NoStop}%
   	\bibitem [{\citenamefont {Kurumaji}\ \emph {et~al.}(2019)\citenamefont
   		{Kurumaji}, \citenamefont {Nakajima}, \citenamefont {Hirschberger},
   		\citenamefont {Kikkawa}, \citenamefont {Yamasaki}, \citenamefont {Sagayama},
   		\citenamefont {Nakao}, \citenamefont {Taguchi}, \citenamefont {hisa Arima},\
   		and\ \citenamefont {Tokura}}]{THE-frustratedmagnet}%
   	\BibitemOpen
   	\bibfield  {author} {\bibinfo {author} {\bibfnamefont {T.}~\bibnamefont
   			{Kurumaji}}, \bibinfo {author} {\bibfnamefont {T.}~\bibnamefont {Nakajima}},
   		\bibinfo {author} {\bibfnamefont {M.}~\bibnamefont {Hirschberger}}, \bibinfo
   		{author} {\bibfnamefont {A.}~\bibnamefont {Kikkawa}}, \bibinfo {author}
   		{\bibfnamefont {Y.}~\bibnamefont {Yamasaki}}, \bibinfo {author}
   		{\bibfnamefont {H.}~\bibnamefont {Sagayama}}, \bibinfo {author}
   		{\bibfnamefont {H.}~\bibnamefont {Nakao}}, \bibinfo {author} {\bibfnamefont
   			{Y.}~\bibnamefont {Taguchi}}, \bibinfo {author} {\bibfnamefont
   			{T.}~\bibnamefont {hisa Arima}},\ and\ \bibinfo {author} {\bibfnamefont
   			{Y.}~\bibnamefont {Tokura}},\ }\href@noop {} {\bibfield  {journal} {\bibinfo
   			{journal} {Science}\ }\textbf {\bibinfo {volume} {365}},\ \bibinfo {pages}
   		{914} (\bibinfo {year} {2019})}\BibitemShut {NoStop}%
   	\bibitem [{\citenamefont {Shiomi}\ \emph {et~al.}(2013)\citenamefont {Shiomi},
   		\citenamefont {Kanazawa}, \citenamefont {Shibata}, \citenamefont {Onose},\
   		and\ \citenamefont {Tokura}}]{TNE-MnGe}%
   	\BibitemOpen
   	\bibfield  {author} {\bibinfo {author} {\bibfnamefont {Y.}~\bibnamefont
   			{Shiomi}}, \bibinfo {author} {\bibfnamefont {N.}~\bibnamefont {Kanazawa}},
   		\bibinfo {author} {\bibfnamefont {K.}~\bibnamefont {Shibata}}, \bibinfo
   		{author} {\bibfnamefont {Y.}~\bibnamefont {Onose}},\ and\ \bibinfo {author}
   		{\bibfnamefont {Y.}~\bibnamefont {Tokura}},\ }\href@noop {} {\bibfield
   		{journal} {\bibinfo  {journal} {Phys. Rev. B}\ }\textbf {\bibinfo {volume}
   			{88}},\ \bibinfo {pages} {064409} (\bibinfo {year} {2013})}\BibitemShut
   	{NoStop}%
   	\bibitem [{\citenamefont {Hirschberger}\ \emph {et~al.}(2020)\citenamefont
   		{Hirschberger}, \citenamefont {Spitz}, \citenamefont {Nomoto}, \citenamefont
   		{Kurumaji}, \citenamefont {Gao}, \citenamefont {Masell}, \citenamefont
   		{Nakajima}, \citenamefont {Kikkawa}, \citenamefont {Yamasaki}, \citenamefont
   		{Sagayama}, \citenamefont {Nakao}, \citenamefont {Taguchi}, \citenamefont
   		{Arita}, \citenamefont {Arima},\ and\ \citenamefont {Tokura}}]{TNE-Gd2PdSi3}%
   	\BibitemOpen
   	\bibfield  {author} {\bibinfo {author} {\bibfnamefont {M.}~\bibnamefont
   			{Hirschberger}}, \bibinfo {author} {\bibfnamefont {L.}~\bibnamefont {Spitz}},
   		\bibinfo {author} {\bibfnamefont {T.}~\bibnamefont {Nomoto}}, \bibinfo
   		{author} {\bibfnamefont {T.}~\bibnamefont {Kurumaji}}, \bibinfo {author}
   		{\bibfnamefont {S.}~\bibnamefont {Gao}}, \bibinfo {author} {\bibfnamefont
   			{J.}~\bibnamefont {Masell}}, \bibinfo {author} {\bibfnamefont
   			{T.}~\bibnamefont {Nakajima}}, \bibinfo {author} {\bibfnamefont
   			{A.}~\bibnamefont {Kikkawa}}, \bibinfo {author} {\bibfnamefont
   			{Y.}~\bibnamefont {Yamasaki}}, \bibinfo {author} {\bibfnamefont
   			{H.}~\bibnamefont {Sagayama}}, \bibinfo {author} {\bibfnamefont
   			{H.}~\bibnamefont {Nakao}}, \bibinfo {author} {\bibfnamefont
   			{Y.}~\bibnamefont {Taguchi}}, \bibinfo {author} {\bibfnamefont
   			{R.}~\bibnamefont {Arita}}, \bibinfo {author} {\bibfnamefont {T.-h.}\
   			\bibnamefont {Arima}},\ and\ \bibinfo {author} {\bibfnamefont
   			{Y.}~\bibnamefont {Tokura}},\ }\href@noop {} {\bibfield  {journal} {\bibinfo
   			{journal} {Phys. Rev. Lett.}\ }\textbf {\bibinfo {volume} {125}},\ \bibinfo
   		{pages} {076602} (\bibinfo {year} {2020})}\BibitemShut {NoStop}%
   	\bibitem [{\citenamefont {Li}\ \emph {et~al.}(2021)\citenamefont {Li},
   		\citenamefont {Wang}, \citenamefont {Wang}, \citenamefont {Yuan},
   		\citenamefont {Song}, \citenamefont {Lou}, \citenamefont {Liu}, \citenamefont
   		{Huang}, \citenamefont {Liu}, \citenamefont {Lei} \emph
   		{et~al.}}]{YMn6Sn6-dirac}%
   	\BibitemOpen
   	\bibfield  {author} {\bibinfo {author} {\bibfnamefont {M.}~\bibnamefont
   			{Li}}, \bibinfo {author} {\bibfnamefont {Q.}~\bibnamefont {Wang}}, \bibinfo
   		{author} {\bibfnamefont {G.}~\bibnamefont {Wang}}, \bibinfo {author}
   		{\bibfnamefont {Z.}~\bibnamefont {Yuan}}, \bibinfo {author} {\bibfnamefont
   			{W.}~\bibnamefont {Song}}, \bibinfo {author} {\bibfnamefont {R.}~\bibnamefont
   			{Lou}}, \bibinfo {author} {\bibfnamefont {Z.}~\bibnamefont {Liu}}, \bibinfo
   		{author} {\bibfnamefont {Y.}~\bibnamefont {Huang}}, \bibinfo {author}
   		{\bibfnamefont {Z.}~\bibnamefont {Liu}}, \bibinfo {author} {\bibfnamefont
   			{H.}~\bibnamefont {Lei}}, \emph {et~al.},\ }\href@noop {} {\bibfield
   		{journal} {\bibinfo  {journal} {Nature communications}\ }\textbf {\bibinfo
   			{volume} {12}},\ \bibinfo {pages} {3129} (\bibinfo {year}
   		{2021})}\BibitemShut {NoStop}%
   	\bibitem [{\citenamefont {Ghimire}\ \emph {et~al.}(2020)\citenamefont
   		{Ghimire}, \citenamefont {Dally}, \citenamefont {Poudel}, \citenamefont
   		{Jones}, \citenamefont {Michel}, \citenamefont {Magar}, \citenamefont
   		{Bleuel}, \citenamefont {McGuire}, \citenamefont {Jiang}, \citenamefont
   		{Mitchell}, \citenamefont {Lynn},\ and\ \citenamefont
   		{Mazin}}]{YMn6Sn6-Nirmal}%
   	\BibitemOpen
   	\bibfield  {author} {\bibinfo {author} {\bibfnamefont {N.~J.}\ \bibnamefont
   			{Ghimire}}, \bibinfo {author} {\bibfnamefont {R.~L.}\ \bibnamefont {Dally}},
   		\bibinfo {author} {\bibfnamefont {L.}~\bibnamefont {Poudel}}, \bibinfo
   		{author} {\bibfnamefont {D.~C.}\ \bibnamefont {Jones}}, \bibinfo {author}
   		{\bibfnamefont {D.}~\bibnamefont {Michel}}, \bibinfo {author} {\bibfnamefont
   			{N.~T.}\ \bibnamefont {Magar}}, \bibinfo {author} {\bibfnamefont
   			{M.}~\bibnamefont {Bleuel}}, \bibinfo {author} {\bibfnamefont {M.~A.}\
   			\bibnamefont {McGuire}}, \bibinfo {author} {\bibfnamefont {J.~S.}\
   			\bibnamefont {Jiang}}, \bibinfo {author} {\bibfnamefont {J.~F.}\ \bibnamefont
   			{Mitchell}}, \bibinfo {author} {\bibfnamefont {J.~W.}\ \bibnamefont {Lynn}},\
   		and\ \bibinfo {author} {\bibfnamefont {I.~I.}\ \bibnamefont {Mazin}},\
   	}\href@noop {} {\bibfield  {journal} {\bibinfo  {journal} {Science Advances}\
   		}\textbf {\bibinfo {volume} {6}},\ \bibinfo {pages} {eabe2680} (\bibinfo
   		{year} {2020})}\BibitemShut {NoStop}%
   	\bibitem [{\citenamefont {Yin}\ \emph {et~al.}(2020)\citenamefont {Yin},
   		\citenamefont {Ma}, \citenamefont {Cochran}, \citenamefont {Xu},
   		\citenamefont {Zhang}, \citenamefont {Tien}, \citenamefont {Shumiya},
   		\citenamefont {Cheng}, \citenamefont {Jiang}, \citenamefont {Lian} \emph
   		{et~al.}}]{TbMn6Sn6-cherngap}%
   	\BibitemOpen
   	\bibfield  {author} {\bibinfo {author} {\bibfnamefont {J.-X.}\ \bibnamefont
   			{Yin}}, \bibinfo {author} {\bibfnamefont {W.}~\bibnamefont {Ma}}, \bibinfo
   		{author} {\bibfnamefont {T.~A.}\ \bibnamefont {Cochran}}, \bibinfo {author}
   		{\bibfnamefont {X.}~\bibnamefont {Xu}}, \bibinfo {author} {\bibfnamefont
   			{S.~S.}\ \bibnamefont {Zhang}}, \bibinfo {author} {\bibfnamefont {H.-J.}\
   			\bibnamefont {Tien}}, \bibinfo {author} {\bibfnamefont {N.}~\bibnamefont
   			{Shumiya}}, \bibinfo {author} {\bibfnamefont {G.}~\bibnamefont {Cheng}},
   		\bibinfo {author} {\bibfnamefont {K.}~\bibnamefont {Jiang}}, \bibinfo
   		{author} {\bibfnamefont {B.}~\bibnamefont {Lian}}, \emph {et~al.},\
   	}\href@noop {} {\bibfield  {journal} {\bibinfo  {journal} {Nature}\ }\textbf
   		{\bibinfo {volume} {583}},\ \bibinfo {pages} {533} (\bibinfo {year}
   		{2020})}\BibitemShut {NoStop}%
   	\bibitem [{\citenamefont {Peng}\ \emph {et~al.}(2021)\citenamefont {Peng},
   		\citenamefont {Han}, \citenamefont {Pokharel}, \citenamefont {Shen},
   		\citenamefont {Li}, \citenamefont {Hashimoto}, \citenamefont {Lu},
   		\citenamefont {Ortiz}, \citenamefont {Luo}, \citenamefont {Li}, \citenamefont
   		{Guo}, \citenamefont {Wang}, \citenamefont {Cui}, \citenamefont {Sun},
   		\citenamefont {Qiao}, \citenamefont {Wilson},\ and\ \citenamefont
   		{He}}]{GaneshPRL}%
   	\BibitemOpen
   	\bibfield  {author} {\bibinfo {author} {\bibfnamefont {S.}~\bibnamefont
   			{Peng}}, \bibinfo {author} {\bibfnamefont {Y.}~\bibnamefont {Han}}, \bibinfo
   		{author} {\bibfnamefont {G.}~\bibnamefont {Pokharel}}, \bibinfo {author}
   		{\bibfnamefont {J.}~\bibnamefont {Shen}}, \bibinfo {author} {\bibfnamefont
   			{Z.}~\bibnamefont {Li}}, \bibinfo {author} {\bibfnamefont {M.}~\bibnamefont
   			{Hashimoto}}, \bibinfo {author} {\bibfnamefont {D.}~\bibnamefont {Lu}},
   		\bibinfo {author} {\bibfnamefont {B.~R.}\ \bibnamefont {Ortiz}}, \bibinfo
   		{author} {\bibfnamefont {Y.}~\bibnamefont {Luo}}, \bibinfo {author}
   		{\bibfnamefont {H.}~\bibnamefont {Li}}, \bibinfo {author} {\bibfnamefont
   			{M.}~\bibnamefont {Guo}}, \bibinfo {author} {\bibfnamefont {B.}~\bibnamefont
   			{Wang}}, \bibinfo {author} {\bibfnamefont {S.}~\bibnamefont {Cui}}, \bibinfo
   		{author} {\bibfnamefont {Z.}~\bibnamefont {Sun}}, \bibinfo {author}
   		{\bibfnamefont {Z.}~\bibnamefont {Qiao}}, \bibinfo {author} {\bibfnamefont
   			{S.~D.}\ \bibnamefont {Wilson}},\ and\ \bibinfo {author} {\bibfnamefont
   			{J.}~\bibnamefont {He}},\ }\href@noop {} {\bibfield  {journal} {\bibinfo
   			{journal} {Phys. Rev. Lett.}\ }\textbf {\bibinfo {volume} {127}},\ \bibinfo
   		{pages} {266401} (\bibinfo {year} {2021})}\BibitemShut {NoStop}%
   	\bibitem [{\citenamefont {Arachchige}\ \emph {et~al.}(2022)\citenamefont
   		{Arachchige}, \citenamefont {Meier}, \citenamefont {Marshall}, \citenamefont
   		{Matsuoka}, \citenamefont {Xue}, \citenamefont {McGuire}, \citenamefont
   		{Hermann}, \citenamefont {Cao},\ and\ \citenamefont {Mandrus}}]{ScV6Sn6}%
   	\BibitemOpen
   	\bibfield  {author} {\bibinfo {author} {\bibfnamefont {H.~W.~S.}\
   			\bibnamefont {Arachchige}}, \bibinfo {author} {\bibfnamefont {W.~R.}\
   			\bibnamefont {Meier}}, \bibinfo {author} {\bibfnamefont {M.}~\bibnamefont
   			{Marshall}}, \bibinfo {author} {\bibfnamefont {T.}~\bibnamefont {Matsuoka}},
   		\bibinfo {author} {\bibfnamefont {R.}~\bibnamefont {Xue}}, \bibinfo {author}
   		{\bibfnamefont {M.~A.}\ \bibnamefont {McGuire}}, \bibinfo {author}
   		{\bibfnamefont {R.~P.}\ \bibnamefont {Hermann}}, \bibinfo {author}
   		{\bibfnamefont {H.}~\bibnamefont {Cao}},\ and\ \bibinfo {author}
   		{\bibfnamefont {D.}~\bibnamefont {Mandrus}},\ }\href@noop {} {\bibfield
   		{journal} {\bibinfo  {journal} {Physical Review Letters}\ }\textbf {\bibinfo
   			{volume} {129}},\ \bibinfo {pages} {216402} (\bibinfo {year}
   		{2022})}\BibitemShut {NoStop}%
   	\bibitem [{\citenamefont {Ma}\ \emph {et~al.}(2021)\citenamefont {Ma},
   		\citenamefont {Xu}, \citenamefont {Yin}, \citenamefont {Yang}, \citenamefont
   		{Zhou}, \citenamefont {Cheng}, \citenamefont {Huang}, \citenamefont {Qu},
   		\citenamefont {Wang}, \citenamefont {Hasan},\ and\ \citenamefont
   		{Jia}}]{QO-RMn6Sn6}%
   	\BibitemOpen
   	\bibfield  {author} {\bibinfo {author} {\bibfnamefont {W.}~\bibnamefont
   			{Ma}}, \bibinfo {author} {\bibfnamefont {X.}~\bibnamefont {Xu}}, \bibinfo
   		{author} {\bibfnamefont {J.-X.}\ \bibnamefont {Yin}}, \bibinfo {author}
   		{\bibfnamefont {H.}~\bibnamefont {Yang}}, \bibinfo {author} {\bibfnamefont
   			{H.}~\bibnamefont {Zhou}}, \bibinfo {author} {\bibfnamefont {Z.-J.}\
   			\bibnamefont {Cheng}}, \bibinfo {author} {\bibfnamefont {Y.}~\bibnamefont
   			{Huang}}, \bibinfo {author} {\bibfnamefont {Z.}~\bibnamefont {Qu}}, \bibinfo
   		{author} {\bibfnamefont {F.}~\bibnamefont {Wang}}, \bibinfo {author}
   		{\bibfnamefont {M.~Z.}\ \bibnamefont {Hasan}},\ and\ \bibinfo {author}
   		{\bibfnamefont {S.}~\bibnamefont {Jia}},\ }\href@noop {} {\bibfield
   		{journal} {\bibinfo  {journal} {Phys. Rev. Lett.}\ }\textbf {\bibinfo
   			{volume} {126}},\ \bibinfo {pages} {246602} (\bibinfo {year}
   		{2021})}\BibitemShut {NoStop}%
   	\bibitem [{\citenamefont {Pokharel}\ \emph {et~al.}(2021)\citenamefont
   		{Pokharel}, \citenamefont {Teicher}, \citenamefont {Ortiz}, \citenamefont
   		{Sarte}, \citenamefont {Wu}, \citenamefont {Peng}, \citenamefont {He},
   		\citenamefont {Seshadri},\ and\ \citenamefont {Wilson}}]{Ganesh-Z2}%
   	\BibitemOpen
   	\bibfield  {author} {\bibinfo {author} {\bibfnamefont {G.}~\bibnamefont
   			{Pokharel}}, \bibinfo {author} {\bibfnamefont {S.~M.~L.}\ \bibnamefont
   			{Teicher}}, \bibinfo {author} {\bibfnamefont {B.~R.}\ \bibnamefont {Ortiz}},
   		\bibinfo {author} {\bibfnamefont {P.~M.}\ \bibnamefont {Sarte}}, \bibinfo
   		{author} {\bibfnamefont {G.}~\bibnamefont {Wu}}, \bibinfo {author}
   		{\bibfnamefont {S.}~\bibnamefont {Peng}}, \bibinfo {author} {\bibfnamefont
   			{J.}~\bibnamefont {He}}, \bibinfo {author} {\bibfnamefont {R.}~\bibnamefont
   			{Seshadri}},\ and\ \bibinfo {author} {\bibfnamefont {S.~D.}\ \bibnamefont
   			{Wilson}},\ }\href@noop {} {\bibfield  {journal} {\bibinfo  {journal} {Phys.
   				Rev. B}\ }\textbf {\bibinfo {volume} {104}},\ \bibinfo {pages} {235139}
   		(\bibinfo {year} {2021})}\BibitemShut {NoStop}%
   	\bibitem [{\citenamefont {Zhang}\ \emph
   		{et~al.}(2022{\natexlab{a}})\citenamefont {Zhang}, \citenamefont {Liu},
   		\citenamefont {Zhang}, \citenamefont {Hou}, \citenamefont {Fu}, \citenamefont
   		{Zhang}, \citenamefont {Gao},\ and\ \citenamefont {Liu}}]{ScMn6Sn6-APL}%
   	\BibitemOpen
   	\bibfield  {author} {\bibinfo {author} {\bibfnamefont {H.}~\bibnamefont
   			{Zhang}}, \bibinfo {author} {\bibfnamefont {C.}~\bibnamefont {Liu}}, \bibinfo
   		{author} {\bibfnamefont {Y.}~\bibnamefont {Zhang}}, \bibinfo {author}
   		{\bibfnamefont {Z.}~\bibnamefont {Hou}}, \bibinfo {author} {\bibfnamefont
   			{X.}~\bibnamefont {Fu}}, \bibinfo {author} {\bibfnamefont {X.}~\bibnamefont
   			{Zhang}}, \bibinfo {author} {\bibfnamefont {X.}~\bibnamefont {Gao}},\ and\
   		\bibinfo {author} {\bibfnamefont {J.}~\bibnamefont {Liu}},\ }\href@noop {}
   	{\bibfield  {journal} {\bibinfo  {journal} {Applied Physics Letters}\
   		}\textbf {\bibinfo {volume} {121}},\ \bibinfo {pages} {202401} (\bibinfo
   		{year} {2022}{\natexlab{a}})}\BibitemShut {NoStop}%
   	\bibitem [{\citenamefont {Roychowdhury}\ \emph {et~al.}()\citenamefont
   		{Roychowdhury}, \citenamefont {Ochs}, \citenamefont {Guin}, \citenamefont
   		{Samanta}, \citenamefont {Noky}, \citenamefont {Shekhar}, \citenamefont
   		{Vergniory}, \citenamefont {Goldberger},\ and\ \citenamefont
   		{Felser}}]{YMn6Sn6_Nernst}%
   	\BibitemOpen
   	\bibfield  {author} {\bibinfo {author} {\bibfnamefont {S.}~\bibnamefont
   			{Roychowdhury}}, \bibinfo {author} {\bibfnamefont {A.~M.}\ \bibnamefont
   			{Ochs}}, \bibinfo {author} {\bibfnamefont {S.~N.}\ \bibnamefont {Guin}},
   		\bibinfo {author} {\bibfnamefont {K.}~\bibnamefont {Samanta}}, \bibinfo
   		{author} {\bibfnamefont {J.}~\bibnamefont {Noky}}, \bibinfo {author}
   		{\bibfnamefont {C.}~\bibnamefont {Shekhar}}, \bibinfo {author} {\bibfnamefont
   			{M.~G.}\ \bibnamefont {Vergniory}}, \bibinfo {author} {\bibfnamefont {J.~E.}\
   			\bibnamefont {Goldberger}},\ and\ \bibinfo {author} {\bibfnamefont
   			{C.}~\bibnamefont {Felser}},\ }\href@noop {} {\bibfield  {journal} {\bibinfo
   			{journal} {Advanced Materials}\ }\textbf {\bibinfo {volume} {34}},\ \bibinfo
   		{pages} {2201350}}\BibitemShut {NoStop}%
   	\bibitem [{\citenamefont {Xu}\ \emph {et~al.}(2022)\citenamefont {Xu},
   		\citenamefont {Yin}, \citenamefont {Ma}, \citenamefont {Tien}, \citenamefont
   		{Qiang}, \citenamefont {Reddy}, \citenamefont {Zhou}, \citenamefont {Shen},
   		\citenamefont {Lu}, \citenamefont {Chang} \emph
   		{et~al.}}]{TbMn6Sn6-2ndthermaltransport}%
   	\BibitemOpen
   	\bibfield  {author} {\bibinfo {author} {\bibfnamefont {X.}~\bibnamefont
   			{Xu}}, \bibinfo {author} {\bibfnamefont {J.-X.}\ \bibnamefont {Yin}},
   		\bibinfo {author} {\bibfnamefont {W.}~\bibnamefont {Ma}}, \bibinfo {author}
   		{\bibfnamefont {H.-J.}\ \bibnamefont {Tien}}, \bibinfo {author}
   		{\bibfnamefont {X.-B.}\ \bibnamefont {Qiang}}, \bibinfo {author}
   		{\bibfnamefont {P.~S.}\ \bibnamefont {Reddy}}, \bibinfo {author}
   		{\bibfnamefont {H.}~\bibnamefont {Zhou}}, \bibinfo {author} {\bibfnamefont
   			{J.}~\bibnamefont {Shen}}, \bibinfo {author} {\bibfnamefont {H.-Z.}\
   			\bibnamefont {Lu}}, \bibinfo {author} {\bibfnamefont {T.-R.}\ \bibnamefont
   			{Chang}}, \emph {et~al.},\ }\href@noop {} {\bibfield  {journal} {\bibinfo
   			{journal} {Nature communications}\ }\textbf {\bibinfo {volume} {13}},\
   		\bibinfo {pages} {1197} (\bibinfo {year} {2022})}\BibitemShut {NoStop}%
   	\bibitem [{\citenamefont {Zhang}\ \emph
   		{et~al.}(2022{\natexlab{b}})\citenamefont {Zhang}, \citenamefont {Koo},
   		\citenamefont {Xu}, \citenamefont {Sretenovic}, \citenamefont {Yan},\ and\
   		\citenamefont {Ke}}]{TbMn6Sn6-thermaltransport}%
   	\BibitemOpen
   	\bibfield  {author} {\bibinfo {author} {\bibfnamefont {H.}~\bibnamefont
   			{Zhang}}, \bibinfo {author} {\bibfnamefont {J.}~\bibnamefont {Koo}}, \bibinfo
   		{author} {\bibfnamefont {C.}~\bibnamefont {Xu}}, \bibinfo {author}
   		{\bibfnamefont {M.}~\bibnamefont {Sretenovic}}, \bibinfo {author}
   		{\bibfnamefont {B.}~\bibnamefont {Yan}},\ and\ \bibinfo {author}
   		{\bibfnamefont {X.}~\bibnamefont {Ke}},\ }\href@noop {} {\bibfield  {journal}
   		{\bibinfo  {journal} {Nature communications}\ }\textbf {\bibinfo {volume}
   			{13}},\ \bibinfo {pages} {1091} (\bibinfo {year}
   		{2022}{\natexlab{b}})}\BibitemShut {NoStop}%
   	\bibitem [{\citenamefont {Canfield}\ \emph {et~al.}(2016)\citenamefont
   		{Canfield}, \citenamefont {Kong}, \citenamefont {Kaluarachchi},\ and\
   		\citenamefont {Jo}}]{canfield}%
   	\BibitemOpen
   	\bibfield  {author} {\bibinfo {author} {\bibfnamefont {P.~C.}\ \bibnamefont
   			{Canfield}}, \bibinfo {author} {\bibfnamefont {T.}~\bibnamefont {Kong}},
   		\bibinfo {author} {\bibfnamefont {U.~S.}\ \bibnamefont {Kaluarachchi}},\ and\
   		\bibinfo {author} {\bibfnamefont {N.~H.}\ \bibnamefont {Jo}},\ }\href@noop {}
   	{\bibfield  {journal} {\bibinfo  {journal} {Philosophical magazine}\ }\textbf
   		{\bibinfo {volume} {96}},\ \bibinfo {pages} {84} (\bibinfo {year}
   		{2016})}\BibitemShut {NoStop}%
   	\bibitem [{\citenamefont {Tritt}(2004)}]{book_thermal}%
   	\BibitemOpen
   	\bibfield  {author} {\bibinfo {author} {\bibfnamefont {T.~M.}\ \bibnamefont
   			{Tritt}},\ }\href@noop {} {\bibfield  {journal} {\bibinfo  {journal}
   			{Properties, and Applications}\ }\textbf {\bibinfo {volume} {2}} (\bibinfo
   		{year} {2004})}\BibitemShut {NoStop}%
   	\bibitem [{\citenamefont {Ashcroft}\ and\ \citenamefont
   		{Mermin}(1976)}]{ashcroft1976solid}%
   	\BibitemOpen
   	\bibfield  {author} {\bibinfo {author} {\bibfnamefont {N.}~\bibnamefont
   			{Ashcroft}}\ and\ \bibinfo {author} {\bibfnamefont {N.}~\bibnamefont
   			{Mermin}},\ }\href@noop {} {\  (\bibinfo {year} {1976})}\BibitemShut
   	{NoStop}%
   	\bibitem [{\citenamefont {Onose}\ \emph {et~al.}(2008)\citenamefont {Onose},
   		\citenamefont {Shiomi},\ and\ \citenamefont
   		{Tokura}}]{Lorenz-number-citation}%
   	\BibitemOpen
   	\bibfield  {author} {\bibinfo {author} {\bibfnamefont {Y.}~\bibnamefont
   			{Onose}}, \bibinfo {author} {\bibfnamefont {Y.}~\bibnamefont {Shiomi}},\ and\
   		\bibinfo {author} {\bibfnamefont {Y.}~\bibnamefont {Tokura}},\ }\href@noop {}
   	{\bibfield  {journal} {\bibinfo  {journal} {Phys. Rev. Lett.}\ }\textbf
   		{\bibinfo {volume} {100}},\ \bibinfo {pages} {016601} (\bibinfo {year}
   		{2008})}\BibitemShut {NoStop}%
   	\bibitem [{\citenamefont {Callen}(1948)}]{Onsager-reltion}%
   	\BibitemOpen
   	\bibfield  {author} {\bibinfo {author} {\bibfnamefont {H.~B.}\ \bibnamefont
   			{Callen}},\ }\href@noop {} {\bibfield  {journal} {\bibinfo  {journal} {Phys.
   				Rev.}\ }\textbf {\bibinfo {volume} {73}},\ \bibinfo {pages} {1349} (\bibinfo
   		{year} {1948})}\BibitemShut {NoStop}%
   	\bibitem [{\citenamefont {Mahan}(1979)}]{mahan1979good}%
   	\BibitemOpen
   	\bibfield  {author} {\bibinfo {author} {\bibfnamefont {G.}~\bibnamefont
   			{Mahan}},\ }\href@noop {} {\bibfield  {journal} {\bibinfo  {journal} {Solid
   				State Phys.}\ }\textbf {\bibinfo {volume} {51}},\ \bibinfo {pages} {81}
   		(\bibinfo {year} {1979})}\BibitemShut {NoStop}%
   	\bibitem [{\citenamefont {Fujishiro}\ \emph {et~al.}(2018)\citenamefont
   		{Fujishiro}, \citenamefont {Kanazawa}, \citenamefont {Shimojima},
   		\citenamefont {Nakamura}, \citenamefont {Ishizaka}, \citenamefont
   		{Koretsune}, \citenamefont {Arita}, \citenamefont {Miyake}, \citenamefont
   		{Mitamura}, \citenamefont {Akiba} \emph {et~al.}}]{MnGe-thermopower}%
   	\BibitemOpen
   	\bibfield  {author} {\bibinfo {author} {\bibfnamefont {Y.}~\bibnamefont
   			{Fujishiro}}, \bibinfo {author} {\bibfnamefont {N.}~\bibnamefont {Kanazawa}},
   		\bibinfo {author} {\bibfnamefont {T.}~\bibnamefont {Shimojima}}, \bibinfo
   		{author} {\bibfnamefont {A.}~\bibnamefont {Nakamura}}, \bibinfo {author}
   		{\bibfnamefont {K.}~\bibnamefont {Ishizaka}}, \bibinfo {author}
   		{\bibfnamefont {T.}~\bibnamefont {Koretsune}}, \bibinfo {author}
   		{\bibfnamefont {R.}~\bibnamefont {Arita}}, \bibinfo {author} {\bibfnamefont
   			{A.}~\bibnamefont {Miyake}}, \bibinfo {author} {\bibfnamefont
   			{H.}~\bibnamefont {Mitamura}}, \bibinfo {author} {\bibfnamefont
   			{K.}~\bibnamefont {Akiba}}, \emph {et~al.},\ }\href@noop {} {\bibfield
   		{journal} {\bibinfo  {journal} {Nature communications}\ }\textbf {\bibinfo
   			{volume} {9}},\ \bibinfo {pages} {408} (\bibinfo {year} {2018})}\BibitemShut
   	{NoStop}%
   	\bibitem [{\citenamefont {Nishino}\ \emph {et~al.}(2006)\citenamefont
   		{Nishino}, \citenamefont {Deguchi},\ and\ \citenamefont
   		{Mizutani}}]{HeuslerhighS}%
   	\BibitemOpen
   	\bibfield  {author} {\bibinfo {author} {\bibfnamefont {Y.}~\bibnamefont
   			{Nishino}}, \bibinfo {author} {\bibfnamefont {S.}~\bibnamefont {Deguchi}},\
   		and\ \bibinfo {author} {\bibfnamefont {U.}~\bibnamefont {Mizutani}},\
   	}\href@noop {} {\bibfield  {journal} {\bibinfo  {journal} {Phys. Rev. B}\
   		}\textbf {\bibinfo {volume} {74}},\ \bibinfo {pages} {115115} (\bibinfo
   		{year} {2006})}\BibitemShut {NoStop}%
   	\bibitem [{\citenamefont {Zhang}\ \emph
   		{et~al.}(2022{\natexlab{c}})\citenamefont {Zhang}, \citenamefont {Xu},
   		\citenamefont {Lee}, \citenamefont {Mao},\ and\ \citenamefont
   		{Ke}}]{MnBi2Te4}%
   	\BibitemOpen
   	\bibfield  {author} {\bibinfo {author} {\bibfnamefont {H.}~\bibnamefont
   			{Zhang}}, \bibinfo {author} {\bibfnamefont {C.~Q.}\ \bibnamefont {Xu}},
   		\bibinfo {author} {\bibfnamefont {S.~H.}\ \bibnamefont {Lee}}, \bibinfo
   		{author} {\bibfnamefont {Z.~Q.}\ \bibnamefont {Mao}},\ and\ \bibinfo {author}
   		{\bibfnamefont {X.}~\bibnamefont {Ke}},\ }\href@noop {} {\bibfield  {journal}
   		{\bibinfo  {journal} {Phys. Rev. B}\ }\textbf {\bibinfo {volume} {105}},\
   		\bibinfo {pages} {184411} (\bibinfo {year} {2022}{\natexlab{c}})}\BibitemShut
   	{NoStop}%
   	\bibitem [{\citenamefont {Owerre}(2017)}]{TTHE-kagomeAFMs}%
   	\BibitemOpen
   	\bibfield  {author} {\bibinfo {author} {\bibfnamefont {S.~A.}\ \bibnamefont
   			{Owerre}},\ }\href@noop {} {\bibfield  {journal} {\bibinfo  {journal} {Phys.
   				Rev. B}\ }\textbf {\bibinfo {volume} {95}},\ \bibinfo {pages} {014422}
   		(\bibinfo {year} {2017})}\BibitemShut {NoStop}%
   	\bibitem [{\citenamefont {Lu}\ \emph {et~al.}(2019)\citenamefont {Lu},
   		\citenamefont {Guo}, \citenamefont {Koval},\ and\ \citenamefont
   		{Jia}}]{TTHE-frustratedAFMs}%
   	\BibitemOpen
   	\bibfield  {author} {\bibinfo {author} {\bibfnamefont {Y.}~\bibnamefont
   			{Lu}}, \bibinfo {author} {\bibfnamefont {X.}~\bibnamefont {Guo}}, \bibinfo
   		{author} {\bibfnamefont {V.}~\bibnamefont {Koval}},\ and\ \bibinfo {author}
   		{\bibfnamefont {C.}~\bibnamefont {Jia}},\ }\href@noop {} {\bibfield
   		{journal} {\bibinfo  {journal} {Phys. Rev. B}\ }\textbf {\bibinfo {volume}
   			{99}},\ \bibinfo {pages} {054409} (\bibinfo {year} {2019})}\BibitemShut
   	{NoStop}%
   	\bibitem [{\citenamefont {van Hoogdalem}\ \emph {et~al.}(2013)\citenamefont
   		{van Hoogdalem}, \citenamefont {Tserkovnyak},\ and\ \citenamefont
   		{Loss}}]{magnetictextureinducedTHE}%
   	\BibitemOpen
   	\bibfield  {author} {\bibinfo {author} {\bibfnamefont {K.~A.}\ \bibnamefont
   			{van Hoogdalem}}, \bibinfo {author} {\bibfnamefont {Y.}~\bibnamefont
   			{Tserkovnyak}},\ and\ \bibinfo {author} {\bibfnamefont {D.}~\bibnamefont
   			{Loss}},\ }\href@noop {} {\bibfield  {journal} {\bibinfo  {journal} {Phys.
   				Rev. B}\ }\textbf {\bibinfo {volume} {87}},\ \bibinfo {pages} {024402}
   		(\bibinfo {year} {2013})}\BibitemShut {NoStop}%
   	\bibitem [{\citenamefont {Iwasaki}\ \emph {et~al.}(2014)\citenamefont
   		{Iwasaki}, \citenamefont {Beekman},\ and\ \citenamefont
   		{Nagaosa}}]{magnondeflectviaskyrmion}%
   	\BibitemOpen
   	\bibfield  {author} {\bibinfo {author} {\bibfnamefont {J.}~\bibnamefont
   			{Iwasaki}}, \bibinfo {author} {\bibfnamefont {A.~J.}\ \bibnamefont
   			{Beekman}},\ and\ \bibinfo {author} {\bibfnamefont {N.}~\bibnamefont
   			{Nagaosa}},\ }\href@noop {} {\bibfield  {journal} {\bibinfo  {journal} {Phys.
   				Rev. B}\ }\textbf {\bibinfo {volume} {89}},\ \bibinfo {pages} {064412}
   		(\bibinfo {year} {2014})}\BibitemShut {NoStop}%
   	\bibitem [{\citenamefont {Ding}\ \emph {et~al.}(2019)\citenamefont {Ding},
   		\citenamefont {Koo}, \citenamefont {Xu}, \citenamefont {Li}, \citenamefont
   		{Lu}, \citenamefont {Zhao}, \citenamefont {Wang}, \citenamefont {Yin},
   		\citenamefont {Lei}, \citenamefont {Yan}, \citenamefont {Zhu},\ and\
   		\citenamefont {Behnia}}]{forthermoelectrictensor}%
   	\BibitemOpen
   	\bibfield  {author} {\bibinfo {author} {\bibfnamefont {L.}~\bibnamefont
   			{Ding}}, \bibinfo {author} {\bibfnamefont {J.}~\bibnamefont {Koo}}, \bibinfo
   		{author} {\bibfnamefont {L.}~\bibnamefont {Xu}}, \bibinfo {author}
   		{\bibfnamefont {X.}~\bibnamefont {Li}}, \bibinfo {author} {\bibfnamefont
   			{X.}~\bibnamefont {Lu}}, \bibinfo {author} {\bibfnamefont {L.}~\bibnamefont
   			{Zhao}}, \bibinfo {author} {\bibfnamefont {Q.}~\bibnamefont {Wang}}, \bibinfo
   		{author} {\bibfnamefont {Q.}~\bibnamefont {Yin}}, \bibinfo {author}
   		{\bibfnamefont {H.}~\bibnamefont {Lei}}, \bibinfo {author} {\bibfnamefont
   			{B.}~\bibnamefont {Yan}}, \bibinfo {author} {\bibfnamefont {Z.}~\bibnamefont
   			{Zhu}},\ and\ \bibinfo {author} {\bibfnamefont {K.}~\bibnamefont {Behnia}},\
   	}\href@noop {} {\bibfield  {journal} {\bibinfo  {journal} {Phys. Rev. X}\
   		}\textbf {\bibinfo {volume} {9}},\ \bibinfo {pages} {041061} (\bibinfo {year}
   		{2019})}\BibitemShut {NoStop}%
   	\bibitem [{\citenamefont {Miyasato}\ \emph {et~al.}(2007)\citenamefont
   		{Miyasato}, \citenamefont {Abe}, \citenamefont {Fujii}, \citenamefont
   		{Asamitsu}, \citenamefont {Onoda}, \citenamefont {Onose}, \citenamefont
   		{Nagaosa},\ and\ \citenamefont {Tokura}}]{thermoelectricandMag}%
   	\BibitemOpen
   	\bibfield  {author} {\bibinfo {author} {\bibfnamefont {T.}~\bibnamefont
   			{Miyasato}}, \bibinfo {author} {\bibfnamefont {N.}~\bibnamefont {Abe}},
   		\bibinfo {author} {\bibfnamefont {T.}~\bibnamefont {Fujii}}, \bibinfo
   		{author} {\bibfnamefont {A.}~\bibnamefont {Asamitsu}}, \bibinfo {author}
   		{\bibfnamefont {S.}~\bibnamefont {Onoda}}, \bibinfo {author} {\bibfnamefont
   			{Y.}~\bibnamefont {Onose}}, \bibinfo {author} {\bibfnamefont
   			{N.}~\bibnamefont {Nagaosa}},\ and\ \bibinfo {author} {\bibfnamefont
   			{Y.}~\bibnamefont {Tokura}},\ }\href@noop {} {\bibfield  {journal} {\bibinfo
   			{journal} {Phys. Rev. Lett.}\ }\textbf {\bibinfo {volume} {99}},\ \bibinfo
   		{pages} {086602} (\bibinfo {year} {2007})}\BibitemShut {NoStop}%
   \end{thebibliography}

\end{document}